\documentclass[11pt,preprint]{aastex}

\newcommand{\kms}{km~s$^{-1}$}
\newcommand{\subsun}{\mbox{$_{\odot}$}}
\newcommand{\etal}{{\it et al.\/}}
\newcommand{\teff}{$T_{\mbox{\scriptsize eff}}$}
\newcommand{\grav}{log($g$)}
\newcommand{\fe}{[Fe/H]}
\newcommand{\mystar}{HE~2148--1247}
\newcommand{\sarastar}{HE~0024--2523}
\newcommand{\compstar}{HE~2344--2800}
\newcommand{\eqw}{$W_{\lambda}$}
\newcommand{\ciso}{$^{12}$C/$^{13}$C}

\begin{document}

\title{Abundance Analysis of \mystar, A Star With Extremely Enhanced 
Neutron Capture Elements\altaffilmark{1}}
\shorttitle{An Extremely Enhanced $n$-Elements Star} 

\author{Judith G. Cohen\altaffilmark{2}, 
Norbert Christlieb\altaffilmark{3,4},
Y.-Z, Qian \altaffilmark{5} \& G.J. Wasserburg \altaffilmark{6} }

\altaffiltext{1}{Based on observations obtained at the
W.M. Keck Observatory, which is operated jointly by the California 
Institute of Technology, the University of California, and the
National Aeronautics and Space Administration.}
\altaffiltext{2}{Palomar Observatory, Mail Stop 105-24,
California Institute of Technology, Pasadena, Ca., 91125}
\altaffiltext{3}{Department of Astronomy and Space Physics, University of
Uppsala, Box 524, SE-75120 Uppsala, Sweden}
\altaffiltext{4}{Marie Curie Fellow, on sabbatical leave from Hamburger
Sternwarte, Germany}
\altaffiltext{5}{School of Physics and Astronomy, University of Minnesota,
Minneapolis, Mn. 55455}
\altaffiltext{6}{Lunatic Asylum, Division of Geological and Planetary
Sciences, California Institute of Technology, Pasadena, Ca., 91125}

\begin{abstract}

Abundances for 27 elements in the very metal poor dwarf star
\mystar\ are presented, including many of the neutron capture
elements. We establish that \mystar\ is a very highly s-process
enhanced star with anomalously high Eu as well, Eu/H $\sim$half
Solar, demonstrating the large addition of heavy 
nuclei at [Fe/H] = $-2.3$ dex. Ba and La are enhanced by a
somewhat larger factor and reach the solar abundance, while Pb
significantly exceeds it, thus demonstrating the addition of
substantial $s$-process material. 
Ba/Eu is ten times the solar $r$-process ratio but much less than that
of the $s$-process, indicating a substantial $r$-process
addition as well.
C and N are also very highly enhanced. We
have found that \mystar\ is a radial velocity variable; it is
probably a small amplitude long period binary. The C, N and the
$s$-process element enhancements thus presumably were produced
through mass transfer from a former AGB binary companion.
The large enhancement of heavy $r$-nuclides also requires an
additional source as this is far above any inventory in the ISM at
such low [Fe/H]. We consider that the $s$-process material was
added by mass transfer of a more massive companion during its
thermally pulsating AGB phase and ending up as a white dwarf. We further hypothesize
that accretion onto the white dwarf from the envelope of the star
caused accretion induced collapse of the white dwarf, forming
a neutron star, which then produced heavy $r$-nuclides and again
contaminated its companion. This mechanism in a binary system can
thus enhance the envelope of the lower mass star in $s$ and 
$r$-process material sequentially.
Through analysis of the neutron capture
element abundances taken from the literature for a large sample of
very metal poor stars,
we demonstrate, as exemplified by \mystar, 
that mass transfer in a suitable binary can be very
efficient in enhancing the heavy elements in a star; it
appears to be capable of enhancing the $s$-process elements
in very metal poor stars to near the solar abundance, but not
substantially above it.
The yield of Pb relative to
Ba appears to vary among very metal poor stars.

\end{abstract}


\section{Introduction}

The most metal deficient stars in the Galaxy provide crucial evidence
on the early epochs of star formation, 
the environments in which various elements
were produced, and the production of elements subsequent to the Big Bang and
prior to contributions from lower mass stars to the ISM. 
The major existing survey for very metal-poor stars is the HK survey
described in detail by Beers, Preston \& Shectman (1985, 1992). 
The stellar inventory of this survey has been scrutinized with
considerable care over the past decade.

In this paper we report the discovery of a peculiar star with highly
enhanced abundances of the heavy neutron capture elements, including
Eu.  \mystar\ was
found in the course of exploiting the Hamburg/ESO Survey (HES)
\citep{wis00,chr01} to find very metal-poor stars. 
We provide a detailed abundance analysis for this star, obtaining
results for 27 elements, including many of the neutron capture elements.
We give a
progress report on our radial velocity monitoring of \mystar,
and conclude with a discussion of the 
implications of its extreme anomalies for galactic nucleosynthesis and chemical
evolution, with emphasis on the issue of the efficiency of mass
transfer in binary systems and a comparison of the observed
abundances in \mystar\ to those predicted under various scenarios.
We present a new version of  accretion induced collapse
in a binary system with heavy $r$-process element production as 
a plausible explanation for the nucleosynthetic
history of \mystar.

We rely heavily in the present work on the procedures and atomic data
for abundance analyses of very metal poor stars
described in our earlier papers reporting the results
of the Keck Pilot Program on Extremely Metal-Poor 
Stars From the HES \citep{cohen02, carretta02,lucatello03}.
At the same time, the spectrum of \mystar,
as well as those of more than 30 additional very metal poor stars
selected from the HES whose spectra we have recently acquired, was
observed with a slight variant of the instrumental
configuration utilized for the Keck Pilot Program.  Thus we 
provide additional details of the procedures
to be applied for the analysis of the
spectrum of \mystar, which will also be used for the
spectra of very metal poor stars now in hand. 

The motivation for studying extremely metal poor stars 
using the HES is reviewed
in \cite{cohen02}.
When our dataset for very metal poor stars
now in hand is analyzed, we will be able
to address in detail the issue of the scatter in abundance ratios
among various elements
from star to star and to
constrain the nucleosynthetic processes operating.  However,
only two of the $\sim$40 high dispersion spectra of newly
discovered very metal poor
stars from the HES that taken thus far at Keck show easily detected 
lines of the key element Eu.   One of these stars, 
\mystar, 
is such an interesting object that we
felt a full analysis should be presented as soon as possible.

\section{Observations \label{section_obs}}

The star \mystar, at coordinates R.A.$(2000.0) =
      21^{h} 51^m 17.8^s$ and $\delta = -12^{\circ} 33' 42''$, was
      selected as a candidate extremely metal-poor star from the
      HES  using automatic
      spectral classification \citep{chr02}. A moderate resolution
      ($\sim2$\,{\AA}) spectrum was taken on November 18, 1999,
      with the 3.5\,m New Technology Telescope at the European
      Southern Observatory, La Silla, Chile.\footnote{ESO Program 
ID 64.L-0005(A).} That spectrum confirmed
      the star to be very metal-poor, and it was therefore
observed at the Keck Telescope using HIRES
\citep{vogt94} on Sep. 28 and 30, 2001.  Six 1200
sec exposures were obtained the first night, and five more
the second, for a total exposure time of 13,200 sec.  The
star was moved slightly along the slit between each exposure.
The observations
were carried out with the slit at the parallactic angle.

A spectral resolution of 45,000
was achieved using a 0.86 arcsec wide slit projecting to 3 pixels in
the HIRES focal plane CCD detector.  As the last order to the
red in the spectra taken for the Keck Pilot Program contained essentially
no new information for such metal poor stars, 
the HIRES configuration was slightly shifted to 
give up the reddest order, thus allowing the inclusion of one more order
to the blue.  We thus cover the region 
from 3805 to 5320\,{\AA} with no gaps to 4910\AA, and small
gaps thereafter to the red as the width of the HIRES CCD
is then smaller than the free spectral range of the echelle.
We will continue with this new configuration until the HIRES
detector is upgraded.

The spectra were exposed such that a minimum SNR of 100 per spectral
resolution element in the continuum at
4500\,{\AA} was achieved.  This SNR calculation utilizes only
Poisson statistics, ignoring issues of cosmic ray removal,
night sky subtraction, flattening, etc.  
(The gain setting of the HIRES
CCD detector is 2.4 e$^-$/ADU; to reach the desired SNR of 100 
per spectral resolution element requires 
1390 ADU/pixel.)
The sum of the HIRES exposures for \mystar\ has 3000 ADU/pixel at
the center of the
echelle order in the continuum near 4500\AA, corresponding to a
signal-to-noise ratio of 145 per spectral resolution element, or
85 per pixel.

This set of HIRES data was reduced
using a combination of Figaro scripts and
the software package MAKEE\footnote{MAKEE was developed
by T.A. Barlow specifically for reduction of Keck HIRES data.  It is
freely available on the world wide web at the
Keck Observatory home page, http://www2.keck.hawaii.edu:3636/.}.
MAKEE automatically applies heliocentric corrections to each of the
individual spectra.
The bias removal, flattening, 
sky subtraction, object extraction and
wavelength solutions with the Th-Ar arc were performed within MAKEE,
after which further processing and analysis was carried out within
Figaro, where the individual spectra were summed.
The continuum fitting to the sum of the individual
spectra (already approximately corrected via the mean signal level
in the flat field spectrum)
uses a 4th-order polynomial to line-free regions of
the spectrum in each order.  The degree of the polynomial was
reduced in orders with H and K of Ca~II or Balmer lines where the
fraction of the order available to define the continuum decreased
significantly.
A scheme of using adjacent orders to help define the polynomial
under such conditions is included in the codes.
The suite of routines for analyzing Ech\`elle
spectra was written by \cite{mccarthy88} within the 
Figaro image processing package
\citep{shortridge93}.

\subsection{Equivalent Widths \label{equiv_widths} }

The search for absorption features present in our HIRES data and the
measurement of their equivalent width (\eqw) was done automatically with
a FORTRAN code, EWDET, developed for a globular cluster project. 
Details of this code and its features are given in \citet{ram01}.
Except in regions affected by CH, CN (or rarely) C$_2$ bands, 
the determination
of the continuum level in these very metal poor stars was easy, 
and the equivalent widths
measured automatically should be quite reliable.  
Except in regions affected by molecular bands, the crowding of lines is
minimal.  Hence we initially use automatic Gaussian fits for \eqw.

The list of lines identified and measured by EWDET is then correlated,
taking the radial velocity into account, 
to the template list of suitable unblended lines 
with atomic parameters described in \S\ref{at_data}
to specifically identify the various atomic lines.   Selected 
wavelength regions (i.e. those
strongly affected by molecular bands, where no individual line
identification can be regarded as secure or the line as unblended)
can be blanked out in this process.

\begin{figure}
\epsscale{1.0}
\plotone{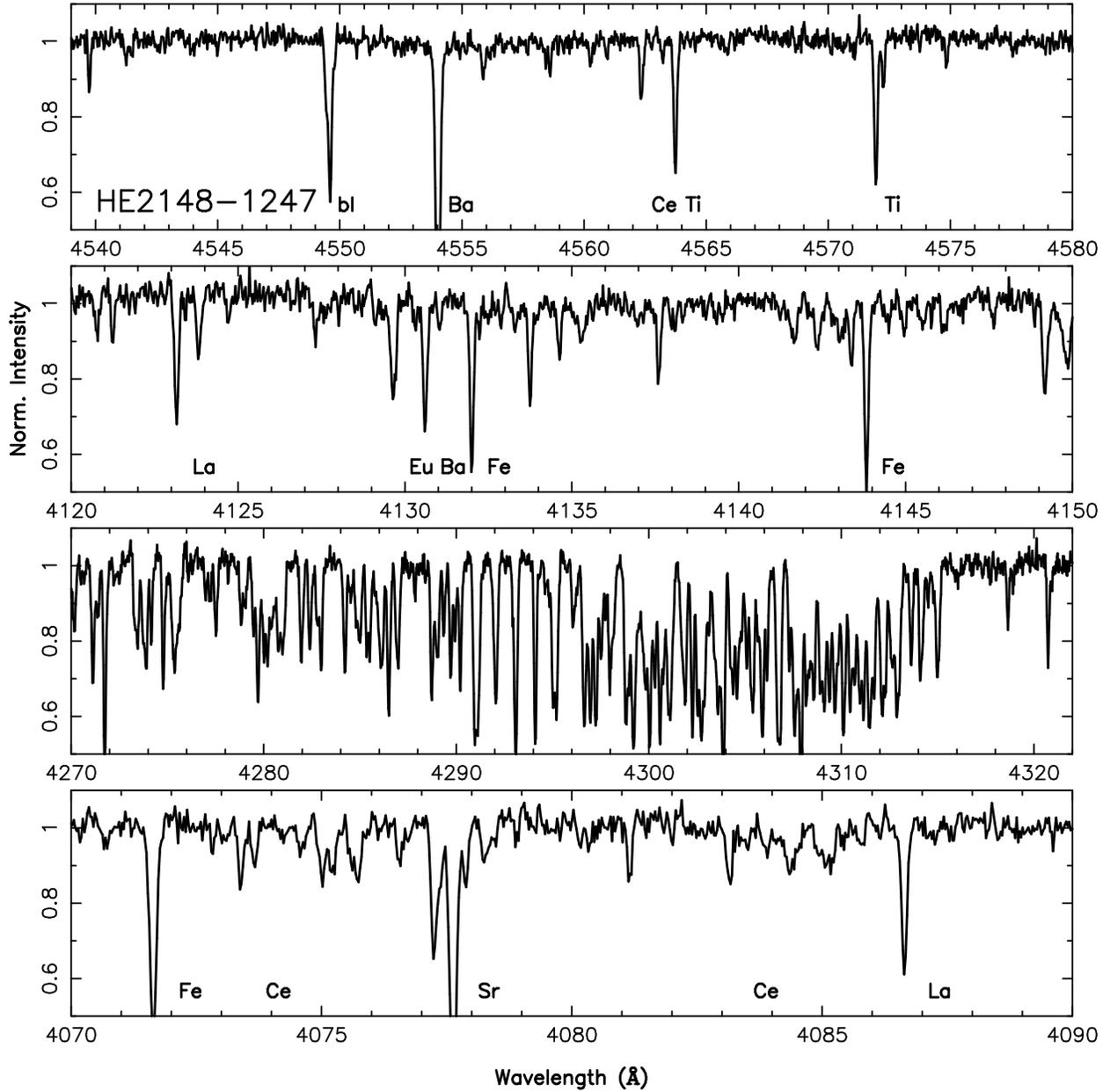}
\caption[]{Extracts from the spectrum of \mystar\ are shown illustrating
the region around the Ba~II line at 4554~\AA, the Eu~II line
at 4129~\AA, the 0-0 bandhead of the G band of CH and the
La~II line at 4086~\AA.   The radial velocity of the star has been removed
by suitable shifts.
Selected strong lines are labelled; the identifications
are displaced 0.4~\AA\ to the red of the line center.  ``Bl''
denotes blends of Fe and Ti lines.
\label{figure_spec}}
\end{figure}

\begin{figure}
\epsscale{0.9}
\plotone{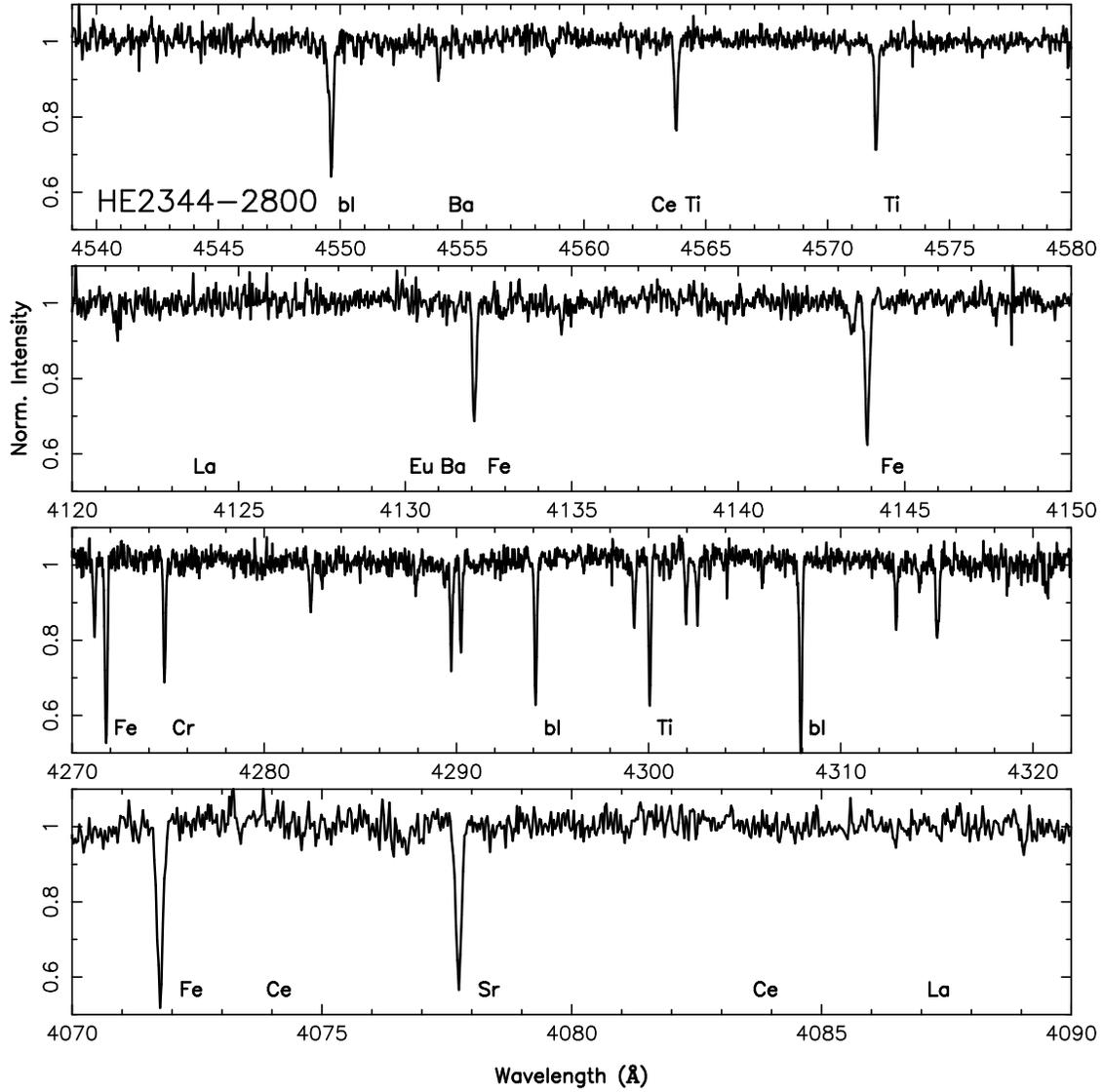}
\caption[]{Extracts from the spectrum of \compstar\ are shown illustrating
the region around the Ba~II line at 4554~\AA, the Eu~II line
at 4129~\AA, the 0-0 bandhead of the G band of CH and the
La~II line at 4086~\AA.  The radial velocity of the star has been removed
by suitable shifts.
The line identifications are those of Figure~\ref{figure_spec}, except
for the region around the CH G bandhead, and
are shown displaced 0.4~\AA\ to the red of the line center.
\label{figure_compstar}}
\end{figure}

At this point we saw the first hint of the extreme peculiarity of \mystar,
as strong Ba~II, La~II, Eu~II,  and Nd~II features turned up in the list
of automatically identified lines.   Selected regions in the
spectrum of \mystar\ 
illustrating some of them are 
shown in Figure~\ref{figure_spec}.  By comparison, \compstar, analyzed in 
\cite{cohen02} and \cite{carretta02}, is
a  ``normal'' very metal poor
main sequence turnoff star
from the HES with about the same \teff, \grav\ and [Fe/H] as \mystar;
it does not show any of the peculiarities of
\mystar\ to be discussed below.  The same spectral regions shown in
Figure~\ref{figure_spec} were extracted from
the spectrum of \compstar\ and are displayed in 
Figure~\ref{figure_compstar} as a comparison.

The automatic identifications were accepted as valid for
lines with \eqw\ $\ge 15$ m\AA.  They were checked by hand
for all lines with smaller \eqw.  The identifications for
all the rare earths in the spectrum of \mystar\ were checked by hand.
The resulting \eqw\ for 159 lines in the spectrum of \mystar\ are 
listed in Table~\ref{table_eqwidth}.

\section{Atomic Data \label{at_data}}

We attempt to assemble a complete list of the unblended atomic features in
this spectral region that might be seen among our entire set of spectra.
We begin with the line list of Carretta \etal\ (2002), created to
support our earlier papers on abundances in extremely metal poor
stars.  With the exceptions noted below, we adopt their $gf$ values  when available.
A few blended lines 
were deleted from their line list.  We then add the unblended lines in the
additional bluest HIRES order.
The transition probabilities for these
were taken from the NIST Atomic Spectra 
Database Version 2.0 (NIST Standard Reference Database \#78),
see \citep{wiese69,mar88,fuh88,wiese96}.
In the Appendix to Carretta \etal\ (2002) we give a comparison of
our adopted transition probabilities for our initial work with
those of NIST.  Very good agreement was noted for all ions
except Mg~I, Ti~II, and Fe~II, where larger dispersions occurred.
The additional lines resulting from the extra blue HIRES order included here
do not contain any lines from these three ions.  We adopted the $gf$ values
of \cite{lawler89} for Sc~II (which agree with those of NIST),
as well as those of \cite{biemont00} for Pb~I.

Since we already have
more than 30 spectra of very metal poor stars 
covering a range of metallicity with [Fe/H] $\le -2$ 
dex\footnote{The 
standard nomenclature is adopted; the abundance of
element $X$ is given by $\epsilon(X) = N(X)/N(H)$ on a scale where
$N(H) = 10^{12}$ H atoms.  Then
[X/H] = log$_{10}$[N(X)/N(H)] $-$ log$_{10}$[N(X)/N(H)]\subsun, and similarly
for [X/Fe].}
taken with the same
instrumental configuration, we also searched for the strongest clean lines
in this spectral region of several ions missing  
from our earlier papers, which we might
hope to detect.  Lines of Ni~I and Co~I were picked up in this way 
in the spectrum of \mystar\ and added,
again with $gf$ values from NIST.
We then added into the template line list the 
strongest expected lines of the rare earths 
using the line list and  $gf$-values from \cite{sneden96}.
This produced a template list of 259 lines, 182 of which arise from
species through atomic number 28 (Ni), with 15 lines from Sr, Y
or Zr, 3 from Ba, and the remainder from the singly ionized rare earths with
the strongest expected lines (La, Ce, Pr, Nd, Sm, Eu, Gd and Dy).

\subsection{Damping Constants}

Values of the collisional broadening parameter $C_6$ 
for van der Waals broadening
are available from detailed fits to solar line
profiles for the Al~I doublet at 3950~\AA\ from 
\cite{bau96}, for three of the five Mg~I lines of interest
from \cite{zhao1998}, and for the 4554~\AA\
Ba~II line (one of the three lines of interest of this ion)
from \cite{mashonkina2000} 
\citep*[chosen in preference to the even
larger value of $C_6$ given for this line in][]{mashonkina1999}.
When such are available, these values are adopted.

Damping constants for the remaining lines of the species
Mg~I, Al~I, Si~I, Ca~I,  Sr~II and Ba~II which might be strong
enough for collisional broadening to be important
are taken from
the recent compilation of collisional damping parameters calculated
from theory by \cite{barklem2000}.
The damping constants for all Fe~I and Fe~II lines were set to twice
that of the Uns\"{o}ld approximation for van der Waals broadening
following \cite{holweger91}. 
The damping constant $C_6$ is set to twice
that of the Uns\"{o}ld approximation for van der Waals broadening
for all other lines, none of which are strong enough to be affected
by the choice of $C_6$.
Note that the above prescription is applied to all spectral lines
irrespective of the value of their measured equivalent width.

\subsection{Hyperfine Structure}

Many of the ions considered here have lines which show appreciable
hyperfine structure.  In some cases, the additional complication of
multiple stable isotopes is also present.   We adopt the
hfs patterns of \cite{mcwilliam98} for Ba~II.  These are calculated
with the solar $r$-process isotope distribution.  We also calculated the
Ba~II hfs patterns from the $r$-process patterns
using the isotopic distributions given
by \cite{arlandini99}.  We use the
hfs and atomic constants of \cite{lawler01a} for La~II and of
\cite{lawler01b} for Eu~II.  We assumed equal abundances for
the two stable isotopes of Eu, $^{151}$Eu and $^{153}$Eu,
which is (approximately) the solar value and that characteristic
of the $r$-process, as discussed in \cite{kappeler89}.
We calculate the hfs patterns for
the relevant Co~I lines from data in \cite{pickering96a}
and \cite{pickering96b}.  The hfs patterns for the relevant
Mn lines were calculated from the data in \cite{booth83}.
There are four lines of Sc~II on
the template list.  For one of these, the hfs pattern was
available from \cite{prochaska00}.  For the other three,
we calculated the hfs patterns using constants from
\cite{arnesen82}, \cite{mansour89} and \cite{villemoes92}.
For the single V~I line in the template list (the only clean strong
line of V in the blue spectral region), which is
not detected in \mystar, we could only
locate hfs parameters for the lower level \citep{childs79}.
We adopted parameters for the upper level comparable to the smallest
splitting found by \cite{childs79} for any of the levels of V~I 
they studied to calculate the hfs pattern for this line.

\section{Solar Abundances \label{section_sun}} 

A number of major advances in abundance analyses have occurred
over the past decade, reflecting progress in
atomic physics and computing power. 
It is now possible to calculate
non-LTE corrections more accurately using more realistic
model atoms.  Furthermore, realistic three dimensional 
radiative-hydrodynamical convection simulations 
have begun to 
become available \citep{asplund00} and to be applied to abundance
determinations in a small number of key situations.
\cite{allende01} and  \cite{allende02} discuss the case of C and of O
in the Sun.

We assume the solar abundances of \cite{anders89},
as updated in \cite{grevesse98}. Those of \cite{anders89} have been derived
using classical 1D abundance analyses of 
carefully selected photospheric lines, with many elements determined from
meteoritic values relative to Si.
\cite{holweger01}
provides updates for the solar abundances of
several key elements including not only
small non-LTE corrections (his maximum $|\Delta(\rm{NLTE})|$ is 0.045 dex)
but also ``granulation'' corrections resulting from the inadequacies of
a classical 1D model given the reality of convection.
His maximum $|\Delta(\rm{gran})|$ is 0.070 dex, for nitrogen,
where the very high excitation of the N~I lines used
in the solar analysis produces
high sensitivity to fluctuations in temperature.

Non-LTE corrections appropriate for low metallicity
stars  are provided in Table~\ref{table_abund} when
available, but we retain the 1D model atmosphere limitation.
Thus, for example, the solar N abundance adopted 
here includes the $-0.07$ dex
granulation correction, but we do not apply it to the
N abundance in \mystar, obtained from a CN band.

We assume that the absolute scale of the adopted
set of $gf$ values is correct over the range of line strength from
the lines we use here for very metal poor stars with high
$gf$ values to the much weaker lines normally used to determine
solar abundances.
\cite{ramirez03} have checked this for a set of lines used to analyze
stars in the Galactic globular cluster
M5, which at [Fe/H] $\sim-1.3$ dex is significantly more metal 
rich than the stars considered
here. They successfully reproduced the standard solar abundance for the
atomic species in common with the present work to within $\pm$0.10 dex
with the exception of Ca~I, which had a discrepancy of 0.16 dex.  

As we use the $gf$-values and the hyperfine structure 
constants of \cite{lawler01a} for La~II, we use their solar
La abundance as well, which is slightly different from that
given by either of the major compilations cited above.
The \cite{lawler01b} deduced solar Eu abundance is in
good agreement with that of \cite{anders89}.  The same situation
holds for Pb~I, where the new $gf$ values of \cite{biemont00}
produce a small shift in the \cite{grevesse98} solar abundance.

The adopted solar abundances are included in the
last column of Table~\ref{table_abund}.

\section{Radial Velocity Monitoring}

After the first spectra revealed the peculiar nature of \mystar, we
began a radial velocity monitoring program, attempting
to observe the star on each of our subsequent Keck/HIRES runs.

The wavelength scale of our spectra are set by observations of a Th-Ar
arc at least twice per night, together with appropriate heliocentric
corrections.
The radial velocity measurement scheme developed for very
metal poor stars in the HES relies upon a set of accurate
laboratory wavelengths for very strong isolated 
features within the wavelength range of
the HIRES spectra.  The wavelengths were taken from the NIST Atomic Spectra 
Database Version 2.0 (NIST Standard Reference Database \#78).
Using an approximate initial $v_r$,
the list of automatically detected lines, restricted to the
strongest detected lines only, generated in the
course of measuring the equivalent widths of the lines 
(see \S\ref{equiv_widths}) in the spectrum of \mystar\
was then searched for
each of these features.  A $v_r$ for each line was determined
from the central wavelength of the best-fit Gaussian, and the
average of these, with a 2.5$\sigma$ clipping reject cycle, 
defined the $v_r$ for the star.
Very metal poor stars which have been observed repeatedly by
George Preston (see Preston \& Sneden 2001)
are used as radial velocity standards.  The
observations of these stars are treated identically as the
sample stars; they serve to monitor the accuracy of the analysis.

Since we are interested in small velocity changes,
we have taken several steps to increase the accuracy of these measurements.
The two laboratory lines between 4270 and 4330 \AA\ are omitted if
the star shows strong CH.  We have also used the spectra
of five bright metal poor stars without strong CH to determine the
small mean offsets between the $v_r$ measured from
each of the laboratory lines.  We set the absolute scale defining
the lab wavelengths of the two unblended lines of the Mg triplet
as 5172.684 and 5183.604\AA.  With that definition, the offsets
range from +4.3 to -2.7 \kms, with only 6 of them exceeding
+1 \kms\ or less than $-1$ \kms. Such differences might arise from
small errors in the laboratory wavelength or repeatable local errors
in the arc fitting procedures.  With these modifications, we are
able to achieve a 1$\sigma$ rms error for the mean $v_r$ of a bright
star of 0.15 \kms, and of $\le0.3$ \kms\ for all but the faintest
of the HES stars for spectra of the quality described 
in \S\ref{section_obs}.

Table~\ref{table_vr} lists the results to date for the two stars
from the HES that we are
currently monitoring, as well as for 
HD~186478, a C-enhanced metal-poor giant star \citep{mcwilliam95}.
(This is one of the stars suggested by G. Preston
as a radial velocity standard.)
In addition to \mystar, we are monitoring HE~0007$-$1832, as when
first observed with HIRES in Sep. 2001, it showed slightly resolved
spectral lines. This star has weaker absorption features than does
\mystar. The errors in the table are the 1$\sigma$ rms error
in the mean velocity, computed assuming Gaussian statistics from the
observed dispersion about the mean and the number of lab lines matched.
With HIRES at Keck, one pixel is 2.0 \kms, and a spectral
resolution element
is $\sim$3 pixels, so a 4 \kms\ change in $v_r$ is equivalent to
a 2 pixel shift in the line center location for each line.

From the data available thus far (three HIRES runs
at roughly 6 month intervals over a year), we find that HE~0007$-$1832
shows no sign of radial velocity changes.  There is some indication
that HD~186478 may have a variable radial 
velocity\footnote{G. Preston advises that his six observations of HD~186478
covered a time span of only 90 days and showed a 1$\sigma$ rms about the mean
of 0.69 \kms.} , while the
radial velocity of \mystar\ is definitely variable and is
probably a small amplitude long period
binary. Radial velocity monitoring will continue for these
three stars.

\section{Stellar Parameters}

The procedure used to derive \teff\ estimates for \mystar\ is fully
explained in \S 4-6 of \cite{cohen02}. Very briefly, \teff\ is derived from
broad-band colors, taking the mean estimates deduced from the de-reddened 
$V-K$ and $V-J$ colors, where the infrared colors are from
2MASS \citep{skrutskie97}.  Measurements on both the 1.5 m
telescope at Palomar Mountain and on the Swope Telescope at the
Las Campanas Observatory yield 
$B = 15.233 \pm0.015$, $V = 14.780 \pm0.023$ mag for \mystar.
We used the grid of predicted broad-band colors and
bolometric corrections of \cite{houdashelt00}, based on the
MARCS stellar atmosphere code \citep{gustafsson75}, and corrected the
colors for reddening by adoption of the extinction maps of 
\cite{schlegel98} to find \teff\ = 6380 K.
This constrains \mystar\ to be either close to the main sequence turnoff, 
or on the horizontal branch. The latter is ruled out by
      the broad, pressure broadened Balmer lines visible
      in the Keck spectra. We therefore adopt the main sequence
      turnoff \grav\ of 3.9 dex, which was derived from the 12 Gyr, 
[Fe/H] = $-2.3$ isochrone of \cite{yi01}.  There is little
sensitivity to the choice of [Fe/H] at this \teff; if
the isochrone for [Fe/H] = $-3.3$ is used instead, then the derived
\grav\ becomes 4.0 dex.
The resulting stellar parameters are \teff=6380 K, 
\grav=3.9 dex and we assumed [Fe/H]=$-$2.5 dex in this step
of our analysis.

\section{Abundance Analysis -- Isolated Lines}

We carried out a classical LTE analysis using the
equivalent widths measured for individual absorption lines.
We employ the grid of stellar atmospheres from \cite{kur93}
(with no overshooting)
and a current version of the MOOG code \citep{sneden73}.
Holding \teff\ and \grav\ fixed, the final overall metallicities [M/H] for
\mystar\ were obtained iteratively, by matching observed \eqw\
with the synthetic ones computed by integrating the equation of
radiative transfer at
different wavelengths along each line for the flux. The
microturbulent velocity, $v_t$, of 1.7 $\pm$0.2 \kms, 
was derived by eliminating any
trend in derived abundances of Fe~I lines with the expected \eqw\ 
\citep*[see ][]{magain84}.
The final abundances are given in Table~\ref{table_abund}.  They
were derived by interpolating
to the desired \teff\ and \grav\ of \mystar\
from the  four nearest model atmospheres in the grid with
[Fe/H]=$-$2.5 dex.

As described above in \S\ref{equiv_widths},
our line list was created automatically (for lines
with \eqw\ $> 15$ m\AA) from the template
list of unblended lines.  Subsequently we examined the residuals 
from the mean abundance
for lines of each element.  Those lines that gave very discrepant results
were checked by hand, and were eliminated if they appeared blended
or there was some other problem.  A relatively small
number of lines were eliminated in this way.

\subsection{Ionization Equilibria and non-LTE \label{section_nonlte}}

We detected two states of ionization for both Fe and Ti
and compare the abundances of the neutral and singly ionized
species in each case.
Since there are a large number
of detected Fe~I lines in principle we 
ignore any Fe~I line with \eqw $> 150$ m\AA.  However, in fact, there are
no Fe~I lines in the spectrum of \mystar\ in this wavelength range
which are that strong (see Table~\ref{table_eqwidth}).
The comparison is excellent in both cases, with log[$\epsilon$(Fe~II)] $-$
log[$\epsilon$(Fe~I)] = $-$0.04 dex (with 10 Fe~II lines detected). For Ti, 
where there are only three weak Ti~I lines detected, this difference is $-$0.02 dex.
This is remarkably good agreement.  Furthermore, the excitation temperature,
which we define here as \teff\ of the model atmosphere for which
the abundance deduced from the set of 58
Fe~I lines is independent of $\chi$, the excitation potential 
of the lower level, is found to be 6480 K, a value which is in 
very good agreement with \teff\ as determined from broad band colors.

This suggests, as does the Fe ionization
equilibrium found for the sample of extremely metal-poor stars
studied by \cite{carretta02}, that non-LTE does not significantly
alter the results of a classical abundance analysis such as presented here.
The theoretical situation
is somewhat unclear, as the results of recent theoretical analyses 
\citep{gra99,the99} disagree on the amplitude to be expected.
\citet{gra99} found that NLTE corrections for Fe lines are very small in
dwarfs of any \teff, and only small corrections ($<$ 0.1 dex) are expected for
stars on the red giant branch.
\citet{the99} found that NLTE corrections become more important as [Fe/H]
decreases, being about 0.2 dex for stars with \fe $\sim -$1.25 dex, and that
ionized lines are not significantly affected by NLTE.  Very recently,
\cite{geh01a} and \cite{geh01b} have carefully calculated the kinetic
equilibrium of Fe, and present in \cite{kor02}
a critique of earlier calculations.  They suggest 
that non-LTE corrections intermediate between
the above sets of values are appropriate 
for Fe~I.

\cite{mashonkina1999} and \cite{mashonkina2000} calculate non-LTE
for Ba~II and Eu~II in low metallicity stars;
corrections appropriate for \mystar\
for these elements from their work are
in the range $-0.1$ to $+0.1$ dex, while \cite{zhao00} find
small non-LTE corrections for Mg~I in very metal poor stars. 
The 3970~\AA\ doublet of
Al~I, the only lines of this species detected in our spectra,
is an exception. The calculated non-LTE 
corrections of \cite{bau97} are large, about +0.6 dex 
for \mystar, while only
+0.05 dex is predicted for the Sun.
The validity of these large non-LTE corrections for this Al~I doublet
has been verified by studies of bright field stars using the resonance
lines as compared to less affected subordinate lines.

On this basis, 
we choose to include non-LTE corrections only for Al~I
in the remainder of this paper.  However, the fourth column
of Table~\ref{table_abund} gives the non-LTE corrections
appropriate to \mystar\ for each species,  
when available, from the sources cited above.

\subsection{Ba and the Rare Earths}

The prominence of lines of Ba and of the rare earths in the spectrum
of \mystar\ is illustrated in Figure~\ref{figure_spec}.
The Ba~II, La~II and Nd~II lines  are stronger than those
of the Sun, even though this star has a very low [Fe/H]
and is somewhat hotter than the Sun, while the Eu~II lines detected are
slightly weaker than the corresponding solar lines.  
The Ba~II lines are the strongest lines in this spectral region, 
excluding only H and K of Ca~II and the Balmer lines.
Ba is a difficult element to analyze, because its lines
are so strong that the choice of damping constants
and hfs become important. However, we
believe, from the comparisons of independent
analyses of slightly more metal rich and slightly cooler
globular cluster giants described in
\cite{ramirez03}, that 
our analysis should be correct to within 0.2 dex.
Use of pure $s$-process isotopic distributions in calculating the hfs
for the Ba~II lines
would decrease the deduced Ba abundance in \mystar\ by 0.08 dex.  

Among the rare earths detected, the abundance of 
Pr is the most uncertain,
as all the lines are very weak (see Table~\ref{table_eqwidth}).
For Sm, the 4318~\AA\ line is
a very solid detection, and is seen in the spectra of \mystar\
taken in Sep. 2002 for radial velocity monitoring as well.
The same holds for the 4103~\AA\ line of Dy~II.
All of the Gd~II lines seen are very weak, but are definitely
real; they too are visible in the Sep 2002 spectra of this star.

\subsection{Sensitivity Study}

The $1\sigma$ rms dispersions for the values of
  log[$\epsilon(X)$] given in Table~\ref{sens_table} contain contributions 
  from various sources of uncertainties 
  in the stellar parameters as well as in the atomic data.  
  The errors in [X/Fe] are typically 
  somewhat smaller than for log($\epsilon$) as the contribution 
  from the uncertainties in the stellar parameters is partially
  cancelled in forming the abundance ratios to Fe; if multiple
  multiple lines of an ion have been detected, that can also lead to
  significantly smaller errors.

Note that the sensitivity to changes in $v_t$ for
those species where hyperfine corrections have
been utilized are small.  However, if no hfs
corrections are used, the sensitivity to changes in $v_t$ is increased
substantially, up to 0.10 dex for a change of 0.2 \kms,
twice the value typical of elements with only strong lines
such as Si~I.

\section{Abundances From Spectral Syntheses}

The abundances of C, N, Th and Pb, as well as the ratio of 
\ciso, were derived from spectral syntheses.  
We generated a line list for each region synthesized by 
adding the strongest atomic lines in the wavelength interval involved to the 
relevant molecular line list.  We use atomic $gf$ values 
from NIST, when available,
or from Kurucz's (1995) atomic line database.  As a first
step in each case, the solar
spectrum was synthesized and compared to that observed 
by \cite{wallace98} with
the FTS at the National Solar Observatory;  this solar spectrum is
available online in digital form.
This was done to verify
that the final line list for each spectral region can reproduce 
the solar spectrum with the nominal solar
abundances.  
Because of the extreme enhancement of the neutron capture elements
in \mystar, the successful completion of a synthesis that matches the
solar spectrum is no guarantee of a line list sufficiently complete
to reproduce the spectrum of \mystar.  In many cases, additional
rare earth lines had to be added to the line list even after
the solar spectrum had been reasonably well matched.
In all the syntheses, the abundances of the various elements with 
lines within that spectral region,
with the exception of the element being tested, were set to those
in Table~\ref{table_abund}.

\subsection{Carbon}

The C abundance was determined from syntheses of the G band of CH.
The molecular line data for CH, including the $gf$ values and the
isotope shifts,
were taken from \cite{jorgensen94} and \cite{jorgensen96}.  
The synthesis was carried out first with an initial guess at \ciso, then
with the value determined from the spectra of \mystar.
However, the main bandhead of the G band at 4305 \AA\ was not used
due to concerns about continuum placement given the strength of the band
and the relatively short length of the echelle orders.  Furthermore
the 0-0 vibrational band is formed higher in the atmosphere, and thus
more subject to any errors in the temperature distribution at high layers,
which is where such are expected to occur as discussed in 
\S\ref{section_highmodel}.
The region from 4318 to 4329 \AA\ was used instead. 

We have no information about the O abundance in \mystar, which is
required to calculate the molecular equilibrium due to the potential
importance of CO. Based
on the characteristics of other heavily C enhanced metal poor stars
\citep*[see, e.g.][]{lucatello03},
we adopt [O/Fe] = +0.5 dex here and throughout this paper. 

\begin{figure}
\epsscale{0.9}
\plotone{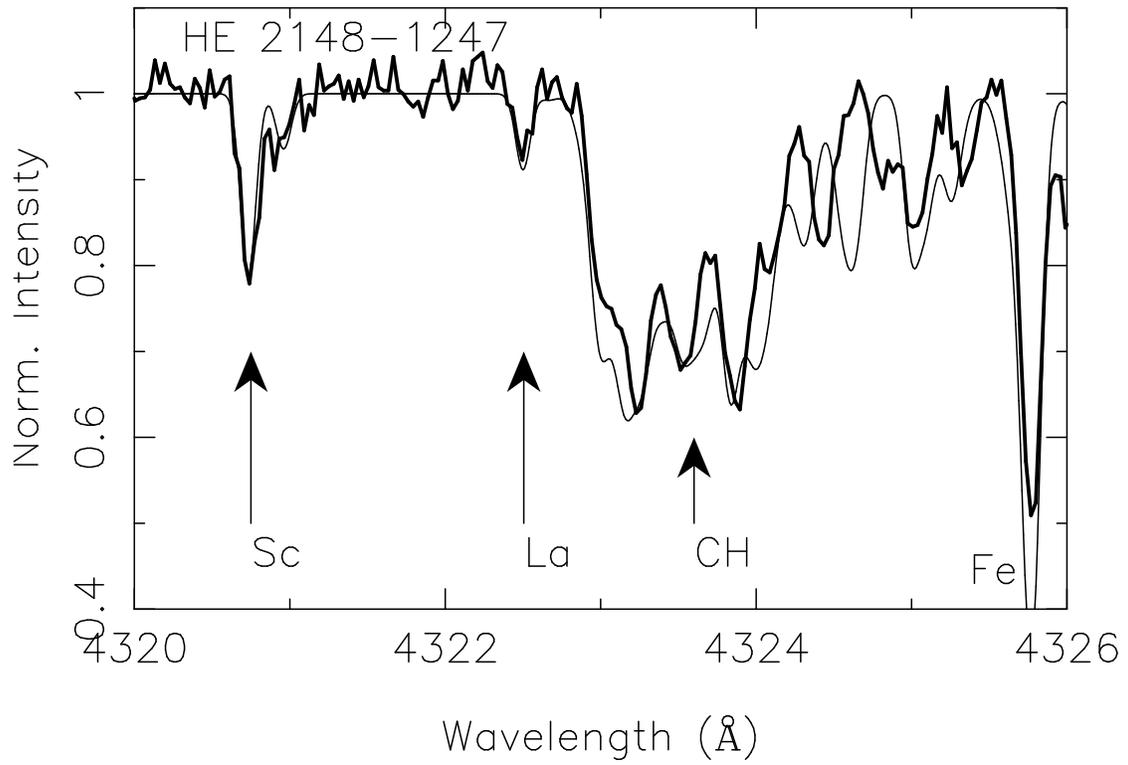}
\caption[]{A synthesis of the region of the CH band from
4320 to 4326~\AA.  The abundances used are those from
Table~\ref{table_abund}.  Note the good fit of the La~II line at
4322.5~\AA.
\label{figure_ch4325}}
\end{figure}

A synthesis
of the region 4218 -- 4329~\AA\ superposed on the observed 
spectrum of \mystar\ is shown in Figure~\ref{figure_ch4325}.  A 
La~II line 4322.5 \AA\ occurs just blueward of the wavelength where
strong CH absorption commences.  The strength of this La~II line
cannot be reproduced without enhancing the La abundance by a factor
comparable to that deduced from the more isolated lines of this
species included in Table~\ref{table_eqwidth}.
The resulting C abundance is high, log($\epsilon$(C)) = $8.08\pm0.2$ dex,
just 0.5 dex below the solar value.

\begin{figure}
\epsscale{1.0}
\plotone{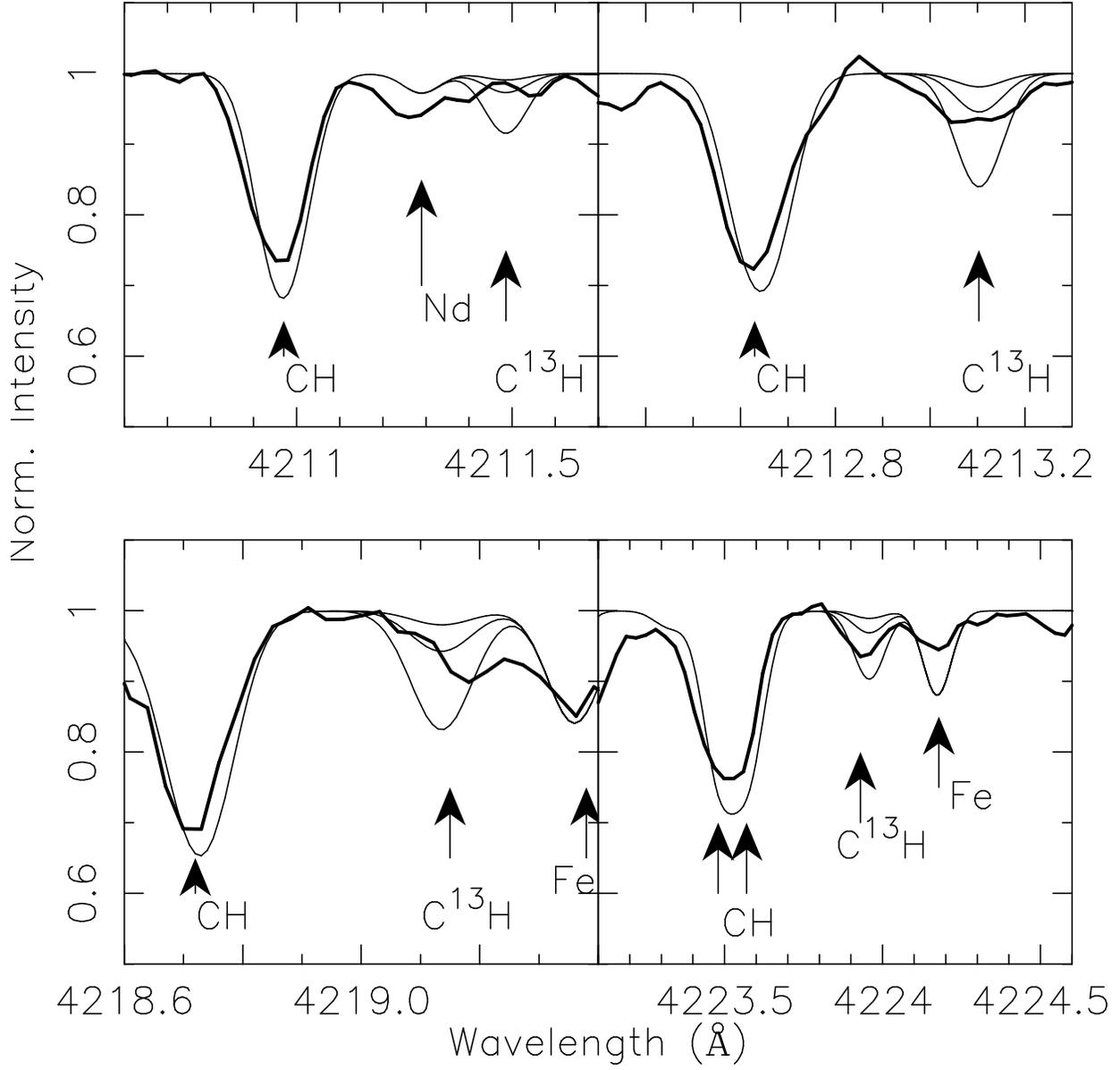}
\caption[]{A synthesis of selected regions with relatively isolated
features of $^{13}$CH overlaid on the spectrum of \mystar. 
Three different isotope ratios are shown, \ciso = 3, 10 and 30.
The abundances used are those from
Table~\ref{table_abund}.  The spectrum in the lower left panel 
is the sum of two adjacent orders.  It has not been smoothed.  
The spectra shown in the other three panels have been slightly smoothed.   
\label{figure_c12c13}}
\end{figure}

The ratio of \ciso\ was determined by synthesizing selected regions
between 4210 and 4225 \AA. 
The wavelengths of the $^{13}$CH lines given by \cite{jorgensen94}
appear to be correct in almost all cases examined in detail.
Figure~\ref{figure_c12c13} shows a spectral synthesis superposed on
the observed spectrum of \mystar\ for various regions with isolated $^{13}$CH
lines in the range 4210 -- 4225~\AA.  Three different isotope ratios are
shown, \ciso\ = 3, 10 and 30.  Examining these features,
we deduce a ratio of \ciso\
for \mystar\ of $\sim$10.  A ratio smaller than 5 can be ruled
out.

\subsection{Nitrogen}

\begin{figure}
\epsscale{0.9}
\plotone{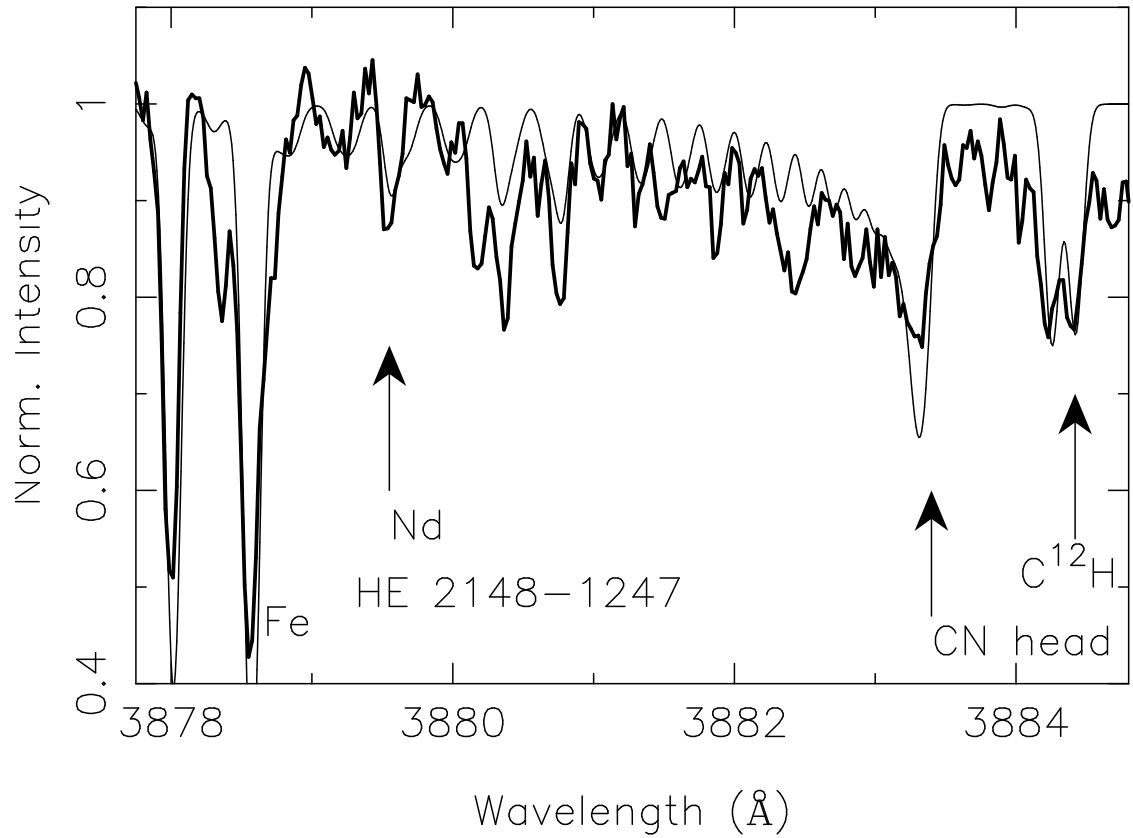}
\caption[]{A synthesis of the region of the bandhead of the
0-0 violet CN band is shown.  The abundances used are those from
Table~\ref{table_abund}.  The wing of the Balmer line H8 at 3889.05~\AA\
lowers the apparent continuum at the red end of the region shown.
\label{figure_cnsyn}}
\end{figure}

The N abundance was determined from the synthesis of a region
around the  0-0  bandhead of the CN violet system shown
in Figure~\ref{figure_cnsyn}. The CN line list
was from \cite{kurucz94}, to which we added the CH lines in that region,
as well as the strongest atomic lines, which are from Fe~I and Cr~I.
We also found we had to add several features of Nd~II.
The high C abundance derived from the syntheses of the CH features
described above was used as an input into the CN synthesis, as was
the ratio derived for
\ciso.  This high C abundance is in fact required to fit the
two strong CH lines 
at 3884.2 and 3884.4~\AA, just redward of the CN bandhead. 
A high N abundance is derived from this synthesis of the 0-0
violet CN band,
whose exact value depends on the choice of C.
To first order, for small changes in C, the required value of N
can be obtained by assuming that the product
of the C and N abundances is constant.

\subsection{Lead and Thorium}

\begin{figure}
\epsscale{1.0}
\plotone{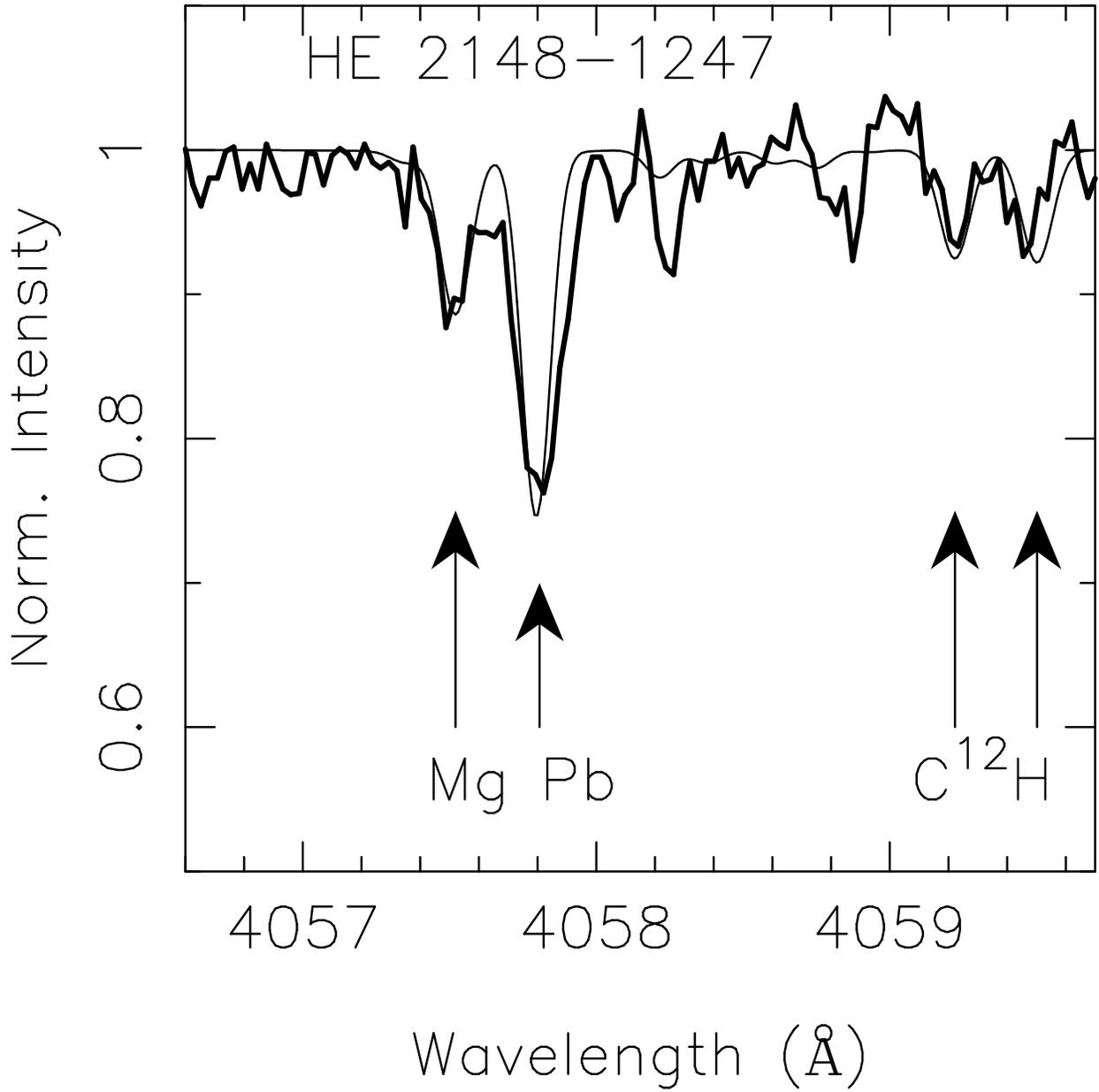}
\caption[]{A spectral synthesis of the region of the Pb~I line at 4057~\AA\
is shown.  The observed spectrom of \mystar\ is shown by the
heavy solid curve, while that of the synthesis is indicated by the thin
curve. The lead abundances used is log[$\epsilon$(Pb)] = +2.65 dex.
\label{figure_lead}}
\end{figure}

The only Pb~I line that is accessible within the wavelength range of
the spectra is at 4057.8~\AA.  This line is extremely strong 
in the spectrum of \mystar.  While 
abundances based on a single line may be considered suspicious, in this 
case the presence of this single strong line of Pb is confirmed beyond 
any doubt as it was
also easily detected in the spectrum  
taken in Sep 2002 as part of the radial velocity monitoring campaign.
As shown
in Figure~\ref{figure_lead}, the Mg~I line at 4057.5~\AA\ 
is a weak feature in
the blue wing of the much stronger Pb~I line.  A spectral synthesis is
not really necessary to establish the Pb abundance, but it was used
to demonstrate that the CH lines in this region
from  $B^2\Sigma^- - X^2\Pi$  transition, while definitely
present, are not playing a significant role, even with the enhanced C abundance
derived for \mystar.  The synthesis was carried out using both 
the hyperfine and isotopic splittings of \cite{simons89} 
assuming a solar distribution among the four stable isotopes of Pb
and also using
a single line to represent this Pb feature.
There is little difference in the resulting Pb abundance, 
$\epsilon(Pb) = +2.8 \pm0.2$ dex.  This is an extremely high abundance
of lead, well above that of the Sun ($\epsilon$(Pb) = 2.0) and
among the highest known for any metal poor star.  

The strongest and cleanest accessible line of thorium in this 
wavelength region is the 4019.139 \AA\ line of Th II.
We adopt log($gf$) = $-0.27$ dex
from \cite{lawler90} for this Th~II line,
but do not adopt their solar Th abundance.
This line is blended with a Co~I line at 4019.125 \AA\
and with a $^{13}$CH line at 4019.136~\AA.
The Ce~II line at 4019.047~\AA, which must be included due to
the enhanced abundances of the rare earths in \mystar, also
contributes to the blue wing of the blend.
(See Morrell, K\"{a}llander \& Butcher 1992 for details;
Johnson \& Bolte 2002 give an up to date line list for
the region of the Th~II 4019~\AA\ line.) 

\begin{figure}
\epsscale{1.0}
\plotone{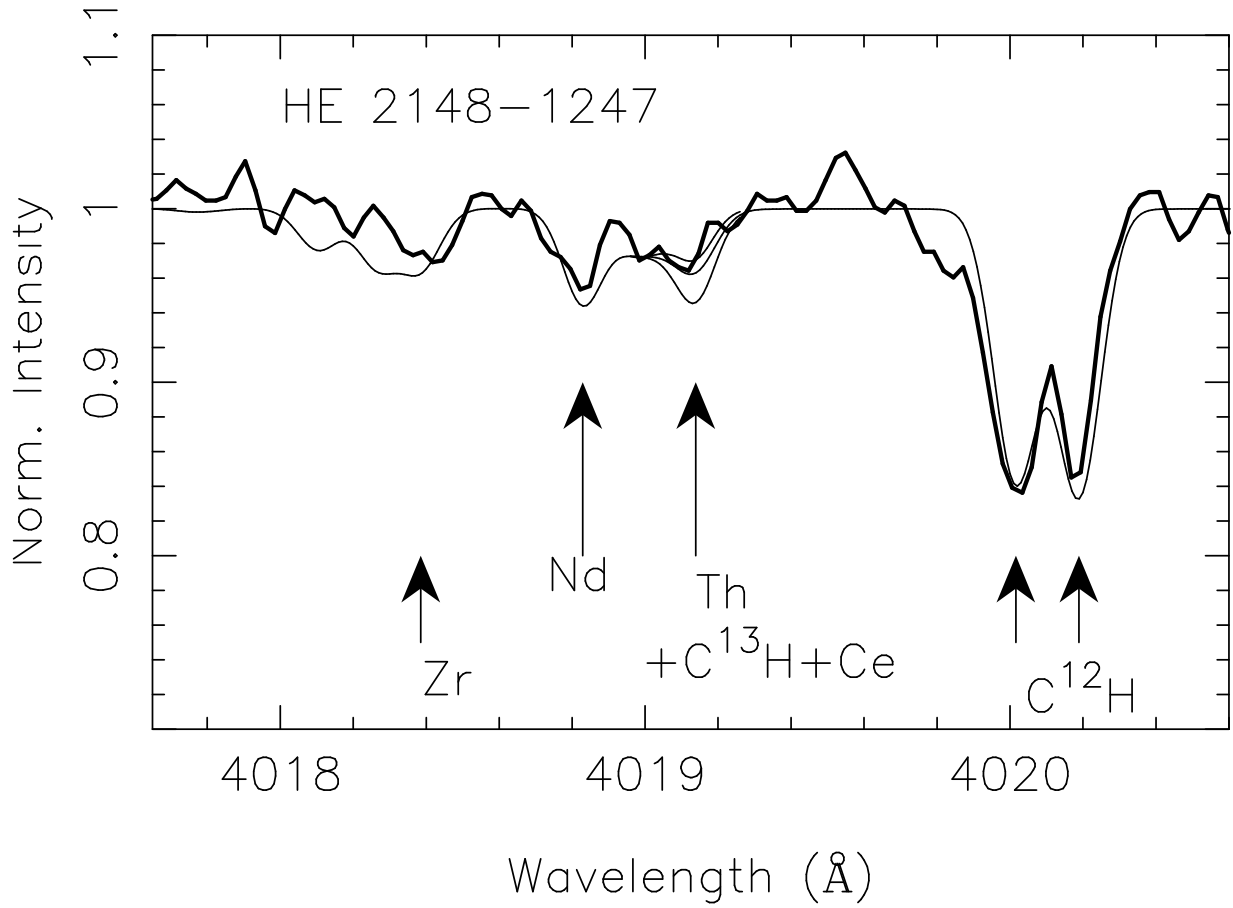}
\caption[]{The slightly smoothed observed spectrum of \mystar\ (shown as
a thick line) is overlaid by the
synthesis of the region around the Th~II line at 4019~\AA\ (thin line).
$\epsilon$(Th) = $-$1.4, $-$0.5 and $-$0.2 dex for the three
cases shown.
The abundances utilized for other elements in the synthesis are 
set to the values  described in Table~\ref{table_abund}.
\label{figure_thorium}}
\end{figure}

Figure~\ref{figure_thorium} shows the synthesis, which used
log[$\epsilon$(Th)] = $-$1.4, $-$0.5 and $-0.2$ dex. 
As expected, a high enhancement of C is required to produce an adequate fit
to the pair of $^{12}$CH lines at 4020.1~\AA.
We infer is a very high Th abundance
log[$\epsilon$(Th)] = $-0.55$ dex.  However, the dominant contributor
to the blend observed in \mystar\ at 4019.1~\AA\ is  $^{13}$CH instead of Th,
and any error in the \ciso\ ratio will strongly affect the derived
Th abundance.  If \ciso\ is less than the value of 10 adopted
in this synthesis,
the Th contribution to the total observed feature at 4019~\AA\ would become 
very small, and thus the deduced Th abundance lowered by a large factor.
The synthesis shows that a line 
at 4019.15~\AA\ is present in the spectrum of \mystar, however
due to the overlapping blends,
the detection of Th is not yet secure. This will require
even higher precision
high resolution spectra. This is unfortunate,
as  a secure detection of Th would make a large
$r$-process contribution mandatory.

The spectrum at the wavelength of the strongest line of U~II
accessible to us
at 3859.57~\AA\ does not have enough precision to provide an interesting
limit on the U abundance.

\subsection{The Effect of the C Excess on Model Atmospheres
\label{section_highmodel}}

Carbon and to a lesser extent nitrogen are highly enhanced in
\mystar, with [C/Fe] = +1.9 dex.  Could such a large enhancement 
of carbon produce excess
absorption of the continuum through various molecular bands,
which would affect the temperature profile in the atmosphere and 
hence the
abundances we derived ?  We first note that such effects will be
larger for individual abundances (i.e. $\epsilon(X)$) than for
the relative abundances of two species. \cite{hill00} compare
the $T-\tau$ profiles for three model atmospheres, one of which
properly includes the C and N enhancements for the stars they
analyze, while another is taken from
the Kurucz grid used here.  The largest difference between
the two above $\tau$(5000\AA)=1 is about 200 K. 
However, the stars they consider,
while having similar C and N enhancements to \mystar, are cool giants
with \teff\ $\sim 4800$ K, much lower than that of \mystar.
As shown by \cite{lucatello03}, the total absorption in the
CH and CN bands for a star similar in [Fe/H] and in [C,N/Fe] to
\mystar\ is not significant when compared to the total flux, approximated
by a blackbody, provided that oxygen is not significantly enhanced
as well.

Strongly enhanced C, especially when accompanied by highly enhanced N, 
does have a significant effect on the ability to detect features
from certain elements whose crucial lines
fall within the wavelength regime covered
by the CH or CN bands.
The Pb~I line at 4057~{\AA}\footnote{This line  is the reddest of the 
expected strong Pb~I lines.  Detection of any of the bluer Pb~I lines
will be even more difficult due to crowding of lines in the spectrum.},
the Th~II line at 4019~\AA\ and
the blue Eu~II lines fall into this category.  It is not always
feasible to obtain red spectra from which to pick up the
6645~\AA\ line of Eu~II, and there are no suitable red lines of any ion of
Pb or Th.  In this context, recall that
\mystar\ is a main sequence turnoff region star, with much weaker
lines and molecular bands because of its higher \teff.

\section{Alternative Explanations}

We reject the notion that the abundance peculiarities seen in
\mystar\ did not originate from a nuclear process, but instead
some type of radiative levitation. Some classes of Ap stars 
display enormous enhancements of the rare earths, the most
extreme known example of which is
HD~101065, Przybylski's star, where the rare earths are 
more or less uniformly enhanced over their
solar abundances by a factor of ${\gtrsim}10^3$. Radiation pressure,
primarily due to absorption lines, can support an atom only
in a very stable atmosphere, and the effect is highly dependent
on luminosity.  Among old and metal-poor populations,
this is seen only in hot horizontal branch stars
(see Behr, Cohen \& McCarthy 2000 and references therein).  
The distribution of the
resulting abundances is that elements with low true abundances rise and
appear enhanced, while those with high initial abundances,
particularly those with few absorption lines, sink and are depleted.
The detailed analysis of Przybylski's star by \cite{cowley00}
yields an abundance distribution for the Fe peak elements which
is not consistent with that typical of metal-poor stars and seen in
\mystar.

Furthermore, as discussed in \S~\ref{section_otherstars},
the abundance peculiarities of \mystar\ are shared by several
other metal poor stars which are cool giants or subgiants whose atmospheres
are kept well mixed by convection.
Radiative levitation is not possible in such stars.

\section{Discussion of Abundances in \mystar \label{section_discuss}}

\begin{figure}
\epsscale{0.9}
\plotone{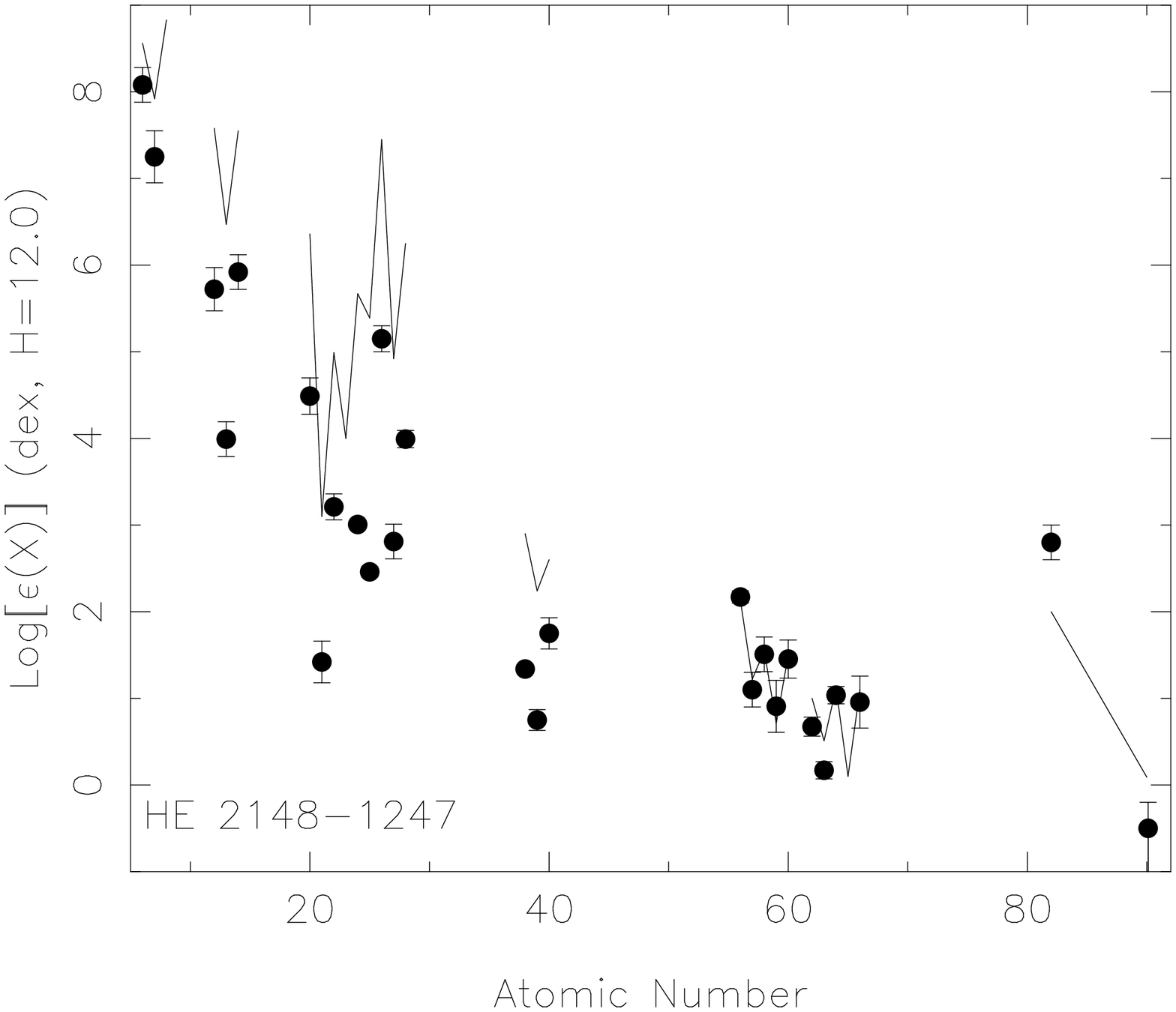}
\caption[]{The abundances of 27 elements in \mystar\ (see 
Table~\ref{table_abund}) are shown
as a function of atomic number, with appropriate error
bars.  The total solar abundances for these elements are indicated by the
line segments.
\label{figure_allelem}}
\end{figure}

\begin{figure}
\epsscale{1.0}
\plotone{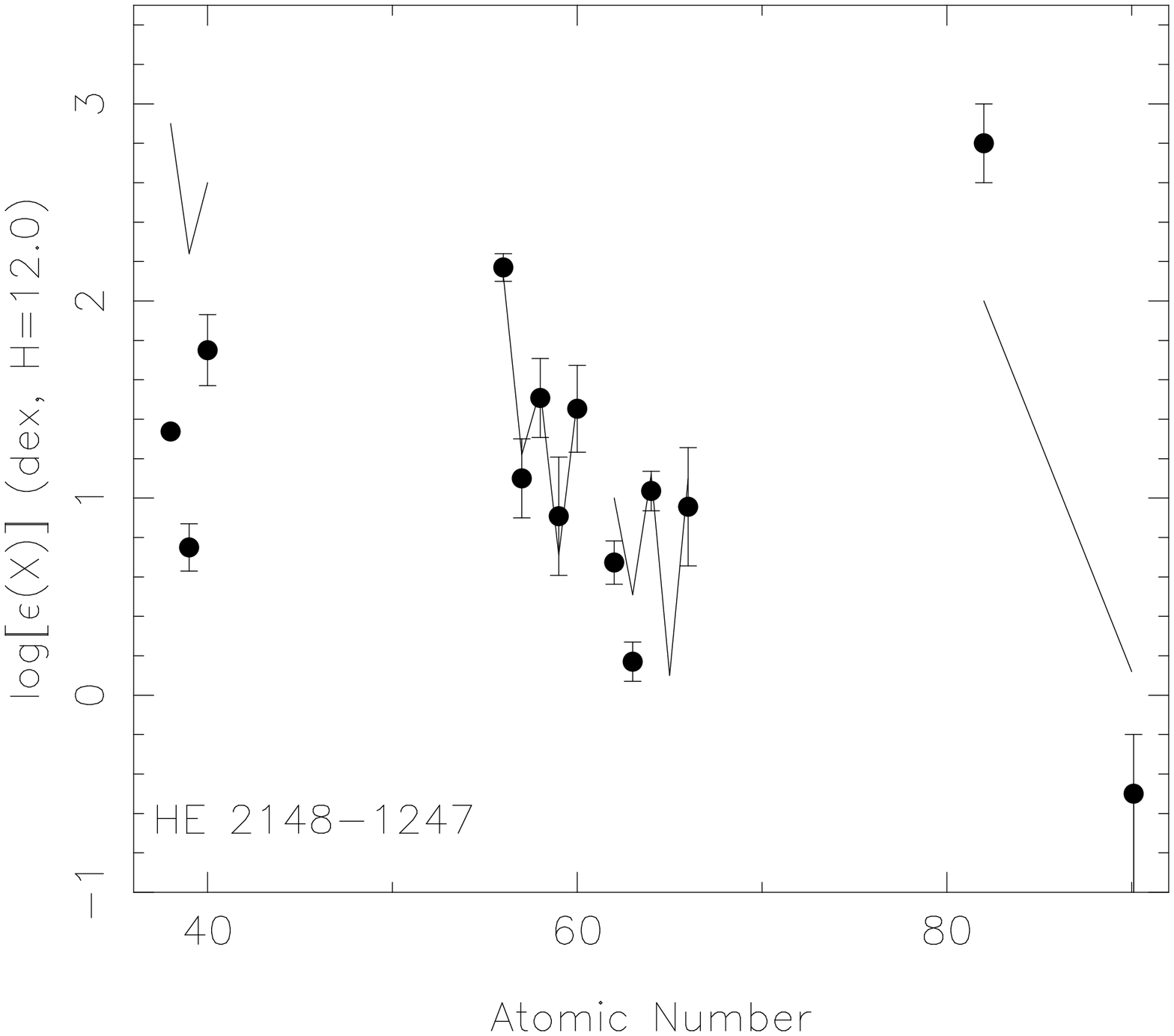}
\caption[]{The abundances of the heavy elements in \mystar\ are shown
as a function of atomic number, with appropriate error
bars.  The total solar abundances for these elements are indicated by the
line segments.
\label{figure_heavy}}
\end{figure}

The elemental abundances determined for \mystar\ are critical for
understanding the sites for the $s$- and $r$-processes of
nucleosynthesis. The relevant observational facts for this star
with a low value of [Fe/H]~$\approx -2.3$ are: (1) the abundances
of the elements from Mg to Ni including Fe relative to hydrogen
are far below the solar values and approximately follow the yield
patterns inferred for very massive stars (VMSs) and some Type II
supernovae [SNe II($L$)] by \cite{qian02}; (2) in general the abundances of
the elements from Ba to Dy relative to hydrogen are remarkably
close to the solar values; (3) Th may be present in high
abundance; (4) the elements Sm, Eu, Gd, and Dy closely follow the
solar $r$-pattern; (5)
the elements Ba, La, Ce, Pr, and Nd approximately follow the solar main
$s$-pattern, demonstrating great $s$-process enrichment; (6) the
Ba/Eu ratio is $\sim 10$ times that for the solar $r$-pattern; and
(7) the Pb/Ba ratio is $\sim 10$ times that for the solar main
$s$-pattern.  In addition, C and N are highly enhanced in \mystar.
Figure~\ref{figure_allelem}  displays the distribution of the
abundances for the 27 elements obtained here. It can be seen that,
except for C and N, almost all elements below $Z = 40$ (Zr) are very
low  compared to solar abundances. Above $Z = 40$ they appear to be
rather close to solar abundances.
Figure~\ref{figure_heavy} focuses on
the region between Sr and Th.
The solid line segments indicate the
total solar abundance for the same set of 
elements\footnote{The solar line also includes the abundances of V and Tb
for clarity.}.  It can be seen that Eu and Sm are distinctly below
the solar curve.

\begin{figure}
\epsscale{1.0}
\plotone{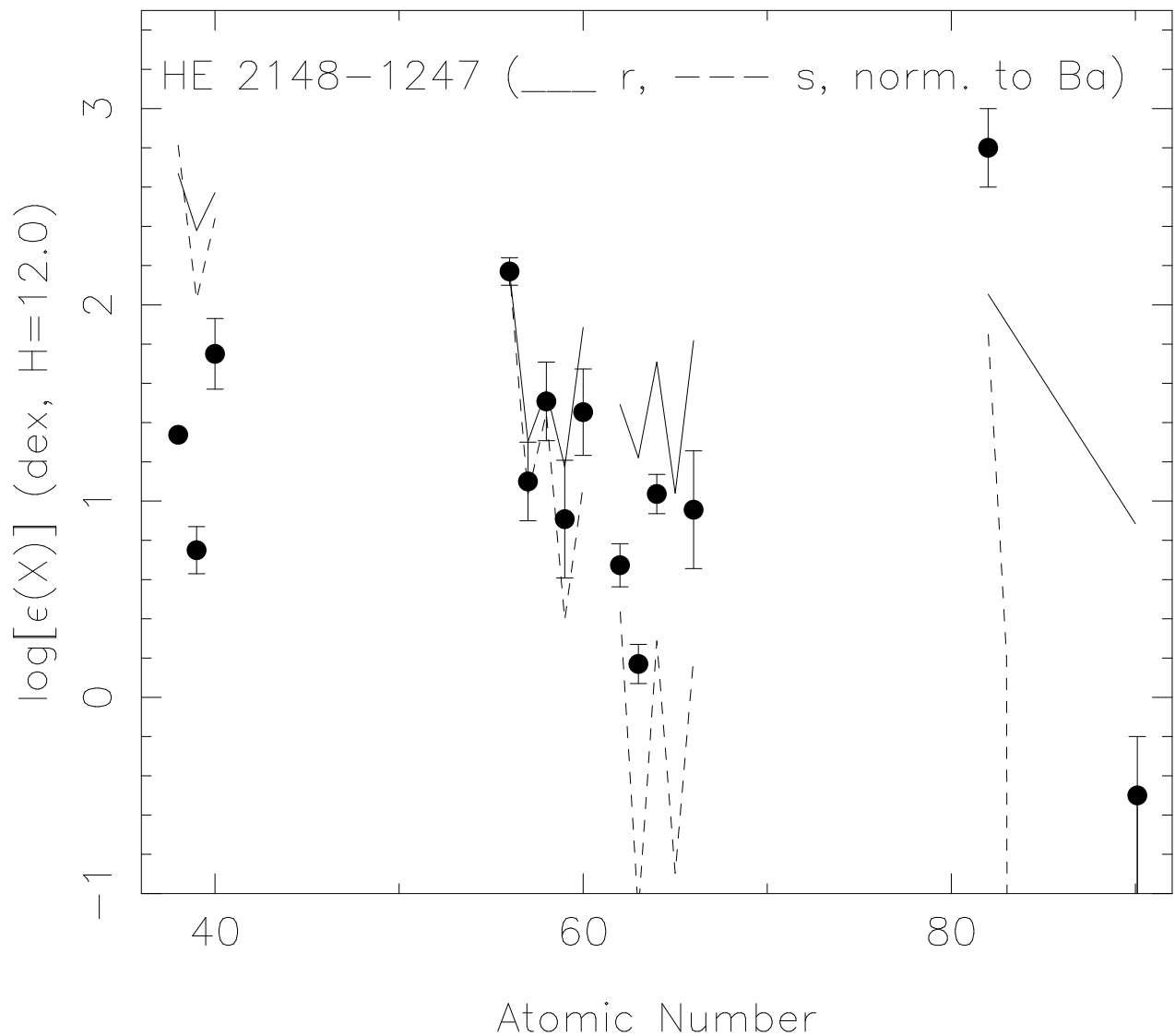}
\caption[]{The derived abundances for \mystar\ for the heavy elements
are shown as a function of atomic number, with appropriate error
bars.  The $r$ and $s$-process solar abundances from \cite{burris00},
scaled to match those of \mystar\ at Ba in each case, are shown
as the solid and dashed line segments respectively.
\label{figure_rsabund}}
\end{figure}

\begin{figure}
\epsscale{1.0}
\plotone{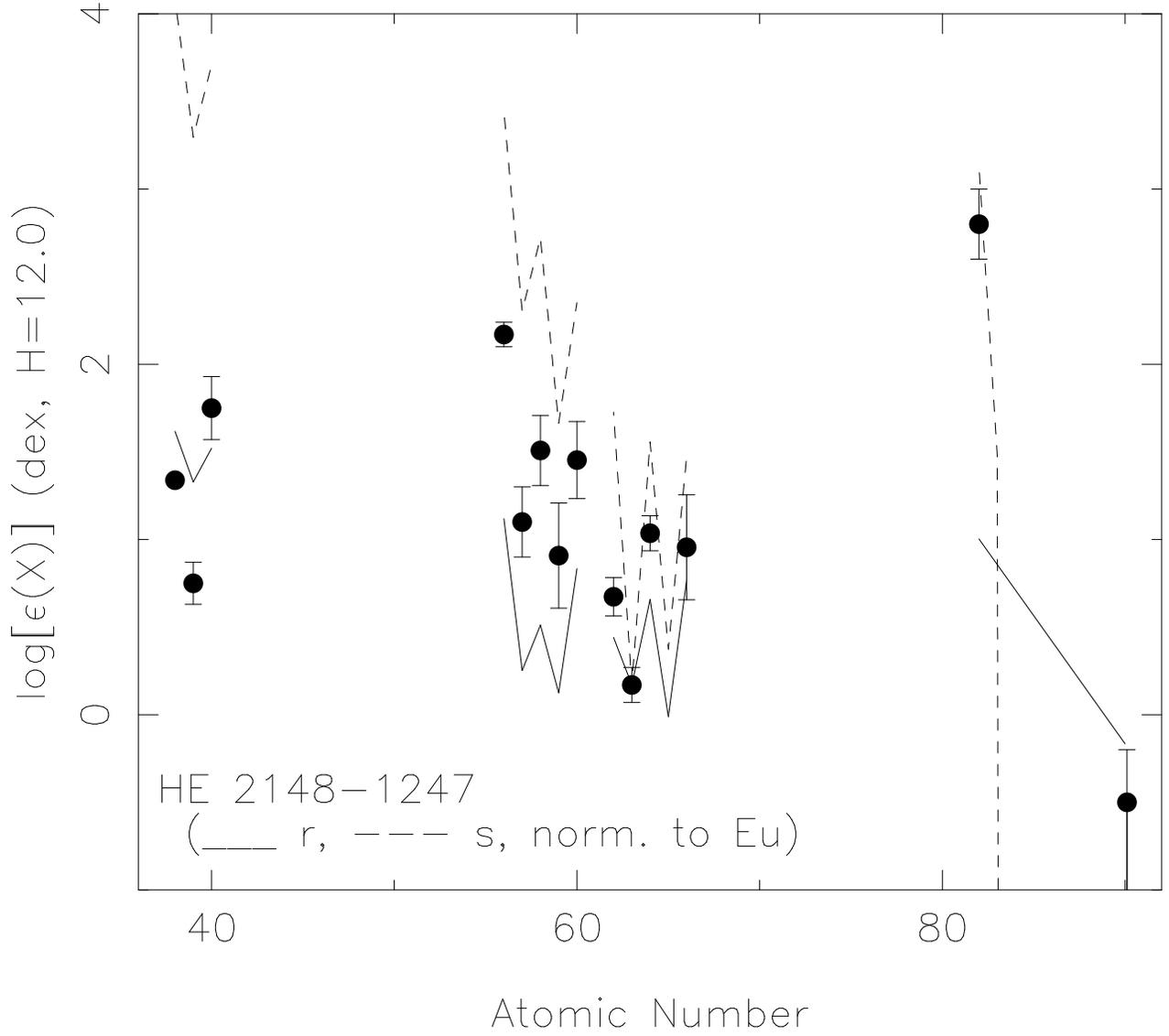}
\caption[]{The same as Figure~\ref{figure_rsabund}, but with the
$r$ and $s$-process solar abundances from \cite{burris00}
scaled to match those of \mystar\ at Eu in each case.
\label{figure_eursabund}}
\end{figure}

If we assume that \mystar\ was initially formed of metal poor gas
with an elemental abundance distribution characteristic of its
present [Fe/H], then it has large enhancements of C and N and
of the heavy elements.    Figure~\ref{figure_rsabund}
shows the abundance distribution from Sr to Th in \mystar\
as in Figure~\ref{figure_heavy} but with the
$r$-process and $s$-process solar abundance distributions,
taken from \cite{burris00}, following
\cite{kappeler89},
indicated by solid and dashed lines.  In both cases, these
are normalized to match the abundance of Ba in \mystar.  
Figure~\ref{figure_eursabund} is similar to 
figure~\ref{figure_rsabund}, but in this case the
$r$-process and $s$-process solar abundance distributions are normalized
to Eu.

Examination of these two figures demonstrates clearly that 
\mystar\ shows large contributions from both the $s$ and the
$r$-process.  Neither component alone is capable of matching the observed heavy
element abundances seen. 
We note that the ratio of Ba to Eu in the solar $r$-process, $s$-process
and total solar inventory \citep*[adopted from][]{burris00} are
is  8.9, 1860 and 59 respectively,
while that of \mystar\ is $\sim$100.  A comparison with La avoids the
difficulties introduced into the abundance analysis 
by the very strong Ba~II lines in \mystar.
The ratio of La to Eu in the solar $r$-process, $s$-process
and total solar inventory \citep*[again adopted from][]{burris00} 
is  1.2, 136 and 4.8 respectively,
while that of \mystar\ is $\sim$8.5.
We have thus attempted to
reproduce the abundance pattern seen in \mystar\ with a sum of
contributions from the two neutron capture processes. We find
that adding 62\% of the total solar $r$-process inventory to
83\% of the total solar $s$-process inventory reproduces the
abundances of all the elements observed in \mystar\ over the range of
atomic number from 56 to 66 to within
the observational errors, as is shown in Table~\ref{table_r_s_sum}.
Thus \mystar, which is, as judged by the elements Na through Zr,
very metal poor, has almost the total solar inventory of heavy neutron capture
elements.  However, Table~\ref{table_r_s_sum} shows that
the observed abundances of Sr, Y and Zr are far (4 to 10 $\sigma$) below 
those predicted by this two component sum.  
Furthermore, the Pb abundance is too high by
more than 4$\sigma$, while the Th abundance, which
has a large observational error, is $\ge$1.3$\sigma$ smaller
than the fit.

\subsection{Comparison with Other Stars \label{section_otherstars}}

Based on examination of the many spectra of very metal poor
stars, we confirm the result of
\cite{nor97} that
enhancement of C is common among very metal poor stars.
Large enhancement of N is somewhat rarer, but not unusual.
We defer discussion of the C and N abundances
to \S\ref{section_implications}.

Our experience to date suggests that
enhancement of the heavy elements is rare among metal
poor stars, and large enhancements are very rare.
Previous examples of $r$-process enhanced very metal poor stars which show
no sign of $s$-process enhancement have been found
by several groups.  The most extreme such stars known
are CS 22892-052 \citep{sne94, sneden96, sne00}
and CS 31082-001 \citep{cay01,hill02}, which have
log[$\epsilon$(Eu)] of $-$0.9 and $-$0.75 dex respectively
with [Fe/H] = $-3.1$ and $-2.9$ dex respectively. 
log($\epsilon$(Eu)) in \mystar\ is much larger, +0.17 dex.
In addition to these two very well studied extremely
$r$-process enhanced stars,
a number of less extreme such stars
have been analyzed by \cite{burris00}, \cite{johnson02} and \cite{cowan02}.
  Among them, the bright star  BD+17$^{\circ}3248$ 
  is of particular interest, because very high precision spectra
  for this star can easily be obtained with HST and from the ground.
  Many r-process elements, not easily seen under normal
  conditions, can thus be detected in the spectrum of this 
  star \citep{cowan02}.

Similarly there are a number of very metal poor stars known to show
large enhancements of $s$-process material. 
There is a sample of 8 such stars in
\cite{aoki01}.  We have also found
another such star in the HES, \sarastar\ \citep{lucatello03},
which shows many similarities to \mystar, is known to be a binary,
but lacks any sign of an $r$-process excess.

\begin{figure}
\epsscale{0.9}
\plotone{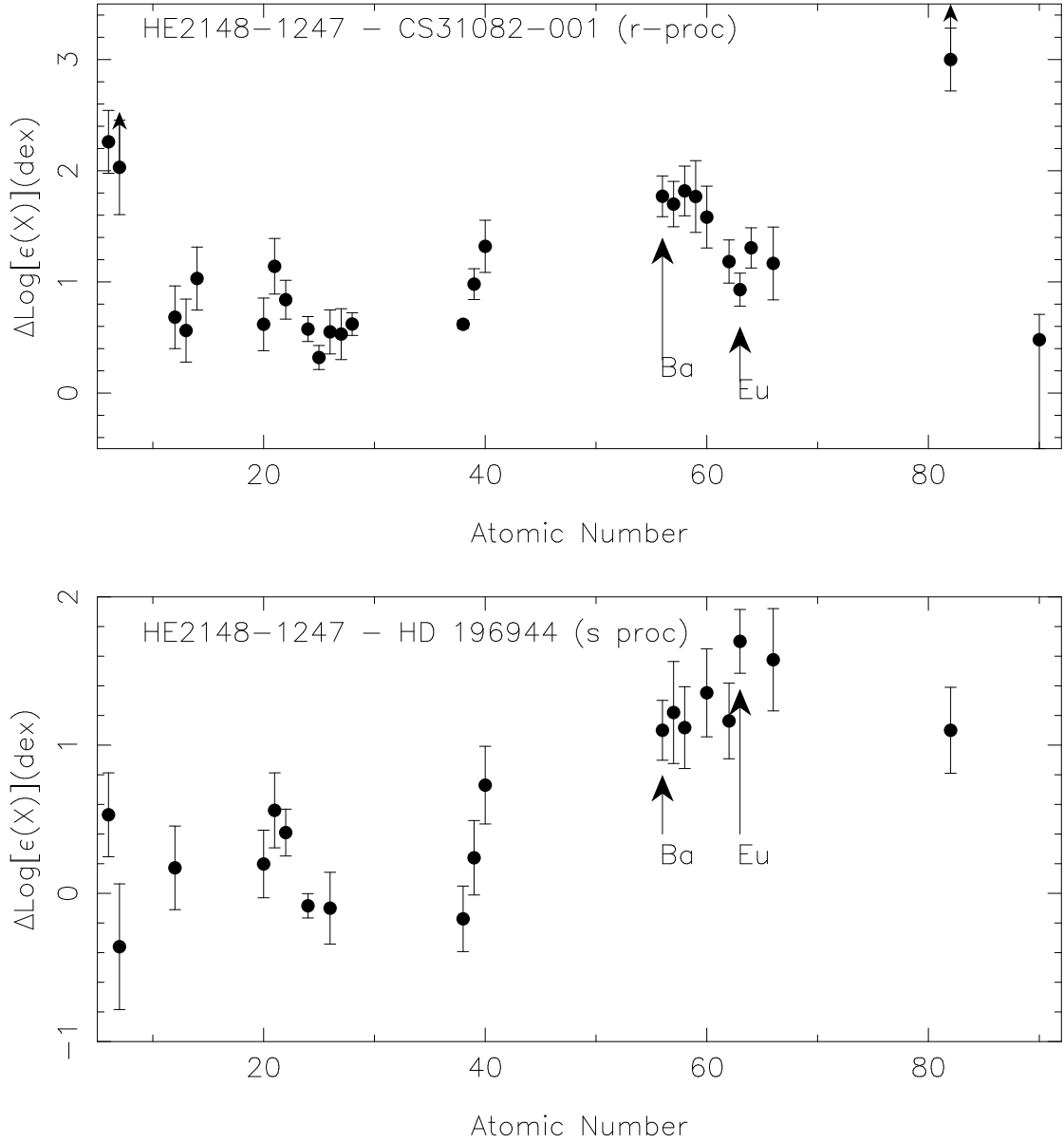}
\caption[]{The abundances in \mystar\ are compared to those
of the $r$-process star CS 31082$-$001 and to the $s$-process
dominated star HD 196944.  Only upper limits for N and for Pb
are available for CS 31082$-$001. The difference in log($\epsilon$) between
the two stars is plotted for each element in common.  
\label{figure_rscomp}}
\end{figure}

Figure~\ref{figure_rscomp} compares the abundance distribution
in \mystar\ to that of the extreme $r$-process star CS 31082$-$001
(data from Hill \etal\ 2002)
and to the $s$-process dominated star HD 196944
(data from Van Eck \etal\ 2001, Aoki \etal\ 2001
and Z\v{a}cs, Nissen \& Schuster 1998).\footnote{A +0.6 dex
correction for non-LTE in Al has been applied throughout.}
The effect of the $s$-process enhancement in the
abundance pattern in \mystar\ is quite obvious in the upper panel
of this figure;
the
enhancement of Ba in the difference between the
abundances in \mystar\ and in CS 31082$-$001 is much larger than that of Eu.
On the other hand, the enhancement of
Eu in \mystar\ relative to that of the $s$-process star is apparent
in the bottom panel, and indicates a strong $r$-process contribution
to the total inventory in \mystar.

\mystar\  is perhaps the best studied example
of a small group of metal poor stars displaying 
signs of both $s$ and $r$-process enhancement.  \cite{hill00}
have studied two additional such stars from the HK Survey.
We attempt to quantify the $r$ and $s$-process contributions in a large
sample of metal poor stars using the Eu abundance to indicate
the former and the Ba abundance to indicate the latter.
We selected a sample of low metallicity ([Fe/H] $< -2.0$ dex)
stars from the literature all of which show signs
of enhanced abundances of the neutron capture elements; all, except one,
have Eu detections.  In addition to the three extremely $r$-process stars
noted above, we include HD~115444, with data from \cite{johnson02}
and from \cite{westin00}.  Less heavily $r$-process enhanced stars
were taken from \cite{johnson02}.  Additional stars were taken
from \cite{hill00}, \cite{nor97} and \cite{aoki03}.
The  $r$ and $s$-process content of each star are plotted in
Figure~\ref{figure_rscontrib}.
The four known stars  with highly enhanced $r$-process abundances
are plotted as the large stars
in the Figure.  They are plotted both at their nominal
abundances (denoted by the large symbols) and 
then with $\epsilon$(Ba)+1$\sigma$
and $\epsilon$(Eu)$-1\sigma$ error, so as to maximize the possible
$s$-process contribution (shown as open triangles in the figure).
\mystar\ is shown by the large filled circle, while
\sarastar, which only has an upper limit for Eu, is shown by
the slightly smaller filled circle.  The solid line denotes the Ba/Eu ratio
expected for pure $r$-process material, the dashed line for 
pure $s$-process material, and the dot-dashed line for material with the
solar ratio of Ba to Eu.  The position of the Sun is marked by the
letter ``S''.

There are two key points here.  The first is the well known wide range
of $r$/$s$ ratio, i.e. the existence of very metal poor stars 
covering the range from those with
neutron capture elements almost purely produced via the $r$-process
to those heavily dominated by the $s$-process.

The second is the magnitude of the contributions.  Note that
Figure~\ref{figure_rscontrib} is plotted with a logarithmic scale
for both axes. All the stars shown
in the figure are very metal poor, and yet in a few of them,
including \mystar,
we see almost the total solar inventory of both $r$ and $s$ process
elements.  On the other hand, the ``pure'' $r$-process stars do not
reach more than 10\% of the solar Eu abundance, nor more
than about 1\% at most of the solar Ba abundance.

\subsection{Implications for the Site of the Nucleosynthesis
\label{section_implications}}

The enhancement of the $r$-process elements in \mystar\ 
as compared to the Fe peak elements is very large,
comparable to or larger than that of the most enhanced pure $r$-process
star discovered to date.  Similarly the
enhancement of the $s$-process elements is also very large, and
again comparable to the most extreme known metal poor stars
in mass transfer binaries.  The enhancement of C and N is also
very large.   Furthermore the
star appears to be a long period binary. 

All scenarios for producing C in low mass stars, whether via the
classical dredge up on the AGB or whether invoking the 
path to mixing at the He flash for very metal poor stars recently
predicted by \cite{fujimoto00}, require that the star be at least
as evolved to incur the He flash.  But \mystar\ is a star near the
main sequence turnoff,
and hence cannot have produced any of the anomalies we see by itself.

\cite{mcclure83} discovered that the CH stars are all single line spectroscopic
binaries. Their spectral peculiarities, including the presence of
strong lines of some of the $s$-process elements, have been attributed 
to mass transfer between the components of a binary system.
If a very metal poor star has as its binary companion a more massive
star that has already evolved to or through the AGB phase, enhancements
of C, N and the $s$-process elements can be produced 
\citep*[see, e.g.][]{jorissen92,han95}.  A binary star model places constraints
on the binary separation and period, as the mass transfer must not
occur while the primary is on the RGB instead of the AGB, and too wide
a separation may make the mass transfer inefficient, as discussed by
\cite{jorissen92}.

\cite{goriely01} and \cite{siess02} discuss the nucleosynthesis
of $s$-elements in low metallicity AGB stars,
and find that even without Fe seed nuclei, extensive $s$-process
production may occur.
Since \mystar\ is a binary, the extremely enhanced C and N
and s-process elements can be explained following the mass transfer
scenario outlined above.
The s-process in very metal-poor stars may
produce copious amounts of Pb
if there is a high neutron to seed ratio \citep{bus01,goriely01}, which is also
observed in \mystar\ and in \sarastar,
another example of an $s$-process enhanced dwarf star 
found in the course of the Keck Pilot Project \citep*[see][]{lucatello03}.

The existence of at least two dwarf stars with substantial C, N 
and $s$-process enhancements suggests that even among the very metal
poor giant stars showing these phenomena, the probable cause is
mass transfer in a binary rather than production via internal nucleosynthesis 
within the star itself.  Recall that very metal poor giants are
assumed to be old
and hence have masses $\sim$0.8$M$\subsun.  Opportunities for internal 
nucleosynthesis and dredge up in such stars are very limited.

\begin{figure}
\epsscale{0.9}
\plotone{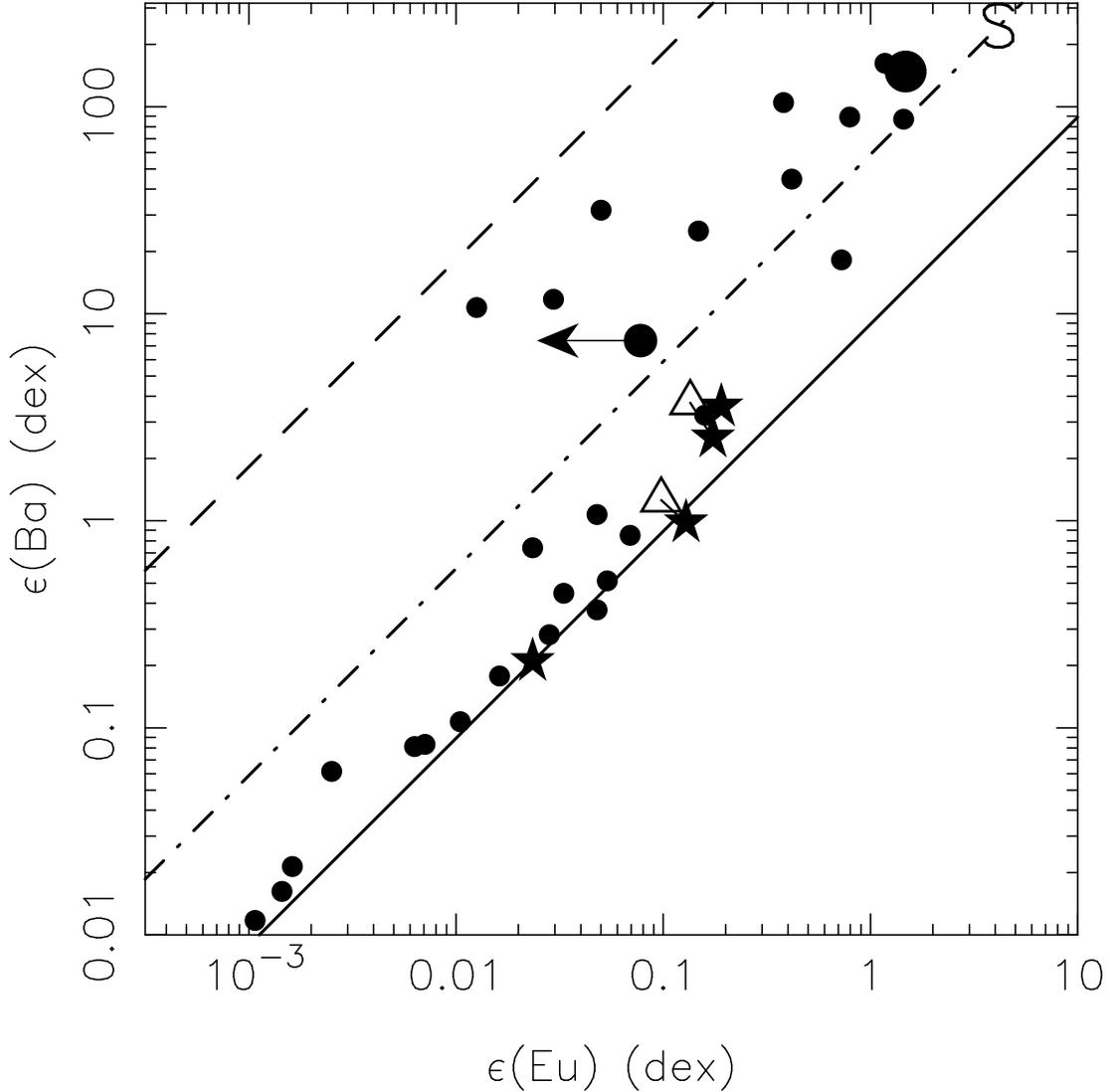}
\caption[]{The ratios of Eu (representing the $r$-process) to Ba 
(representing the $s$-process) are shown for
a sample of metal poor stars, all with [Fe/H] $< -2.0$ dex.  The sources
for the abundances are given in the text.
The large stars denote the four highly enhanced $r$-process stars
with their nominal abundances.  In two cases, the results are shown 
for the abundances including the errors so as to maximize the 
resulting $s$-process contribution.  These are indicated by 
open triangles connected by a line to the symbol denoting the nominal
abundance case. 
\mystar\ is shown as the large
filled circle.  \sarastar\ is shown as a slightly smaller filled circle
with an arrow indicating only an upper limit for $\epsilon$(Eu).
The solid line denotes the $r$-process Ba/Eu ratio, the dashed line
denotes that for the $s$-process, and the dot-dashed line denotes
the Solar Ba/Eu ratio.  The letter ``S'' indicates the position of the Sun.
\label{figure_rscontrib}}
\end{figure}

\begin{figure}
\epsscale{0.9}
\plotone{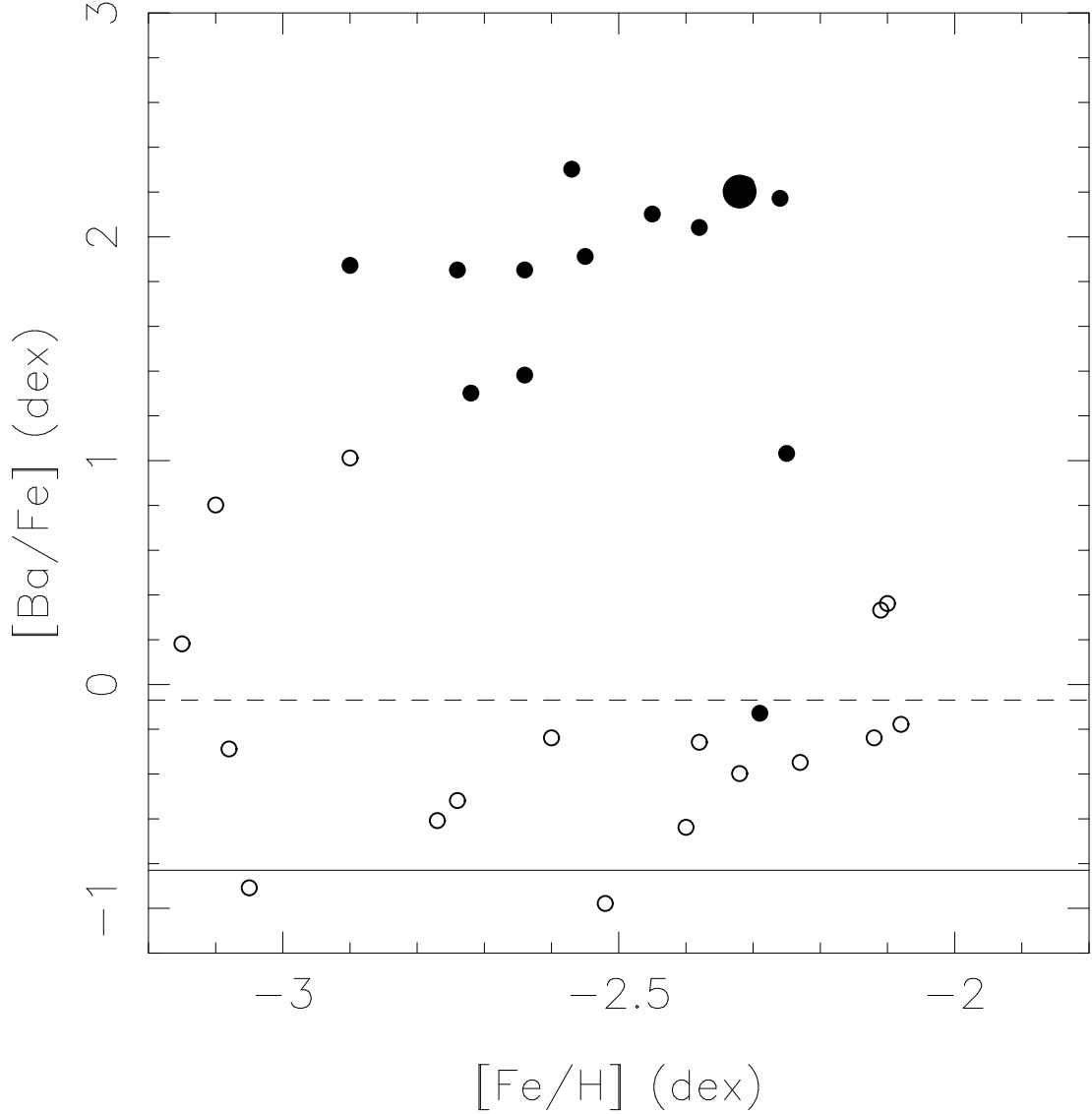}
\caption[]{The enhancement of the $s$-process elements
parameterized as [Ba/Fe]
over the nominal $s$-process inventory expected
for the [Fe/H] of the star is plotted as a function
of [Fe/H] for the sample of very metal poor stars 
shown in Figure~\ref{figure_rscontrib}. 
Those stars with Ba/Eu $< 25$ (i.e. the $r$-dominated stars) are shown
as open circles, while stars with Ba/Eu $> 25$ 
(i.e. the $s$-dominated stars) are shown as filled circles.  \mystar\
is shown with a larger symbol.  The solid horizontal line
indicates the fraction of Ba in the Sun produced by the $r$-process; 
the dashed line shows that fraction for the solar $s$-process.  
\label{figure_senhance}}
\end{figure}

Figure~\ref{figure_senhance} illustrates the
$s$-process enhancement parameterized as [Ba/Fe] 
as a function of [Fe/H] for the same sample of very metal poor stars
shown in Figure~\ref{figure_rscontrib}.  Those stars with
Ba/Eu $< 25$ (i.e. those which are $r$-process dominated) are shown
as open circles, while those with Ba/Eu $> 25$ are shown as filled circles.  
Recall that all these stars have [Fe/H] $\le -2.0$ dex.
Note the very high $s$-process enhancement (a factor of 200
over that expected from the solar $s$-process inventory of Ba
for the star's [Fe/H] value) achieved by the
most extreme of the $s$-process dominated stars.  This large Ba
enhancement is achieved at [Fe/H] ${\sim}-$2.5 dex.
Comparing this, and the Pb enhancement of \mystar,
with Figure~12 of \cite{busso99} suggests that some adjustments
are required in the model of galactic nucleosynthesis from AGB stars.
Furthermore, these excesses in \mystar\ must be confined 
mostly to the stellar surface.  There
cannot have been much sedimentation or mixing of the additional
material into the interior of \mystar, otherwise
the amount of additional neutron capture material required would become
prohibitively large.  

Figure~\ref{figure_rscontrib} suggests that the efficiency for 
the complex chain of events 
proposed to explain highly $s$-process enhanced very metal poor
stars, which includes nucleosynthesis
in AGB stars, dredge up to the surface of the AGB star, 
and mass transfer across a binary, can be surprisingly efficient;
we have effectively taken a very low metallicity
star and given it a substantial fraction of the solar $s$-process
abundances.  This is a much more effective way of enhancing the
abundances  for a particular
star in a binary system with a suitable mass ratio
than is the production and distribution into the ISM of
a single Type II SN.
We speculate that {\it {all}} of the very metal poor stars with more
than 10\% of the solar inventory of neutron capture elements
must be (or have been) binaries.  At least a few of the known 
extremely $s$-process enhanced stars have been found to be binaries,
in particular the two from the HES (\mystar, \sarastar\ Lucatello
\etal\ 2003), LP 625$-$44 \citep{aoki00,aoki02}, as well as CS 29526$-$110
\citep{aoki03}. 

\begin{figure}
\epsscale{0.9}
\plotone{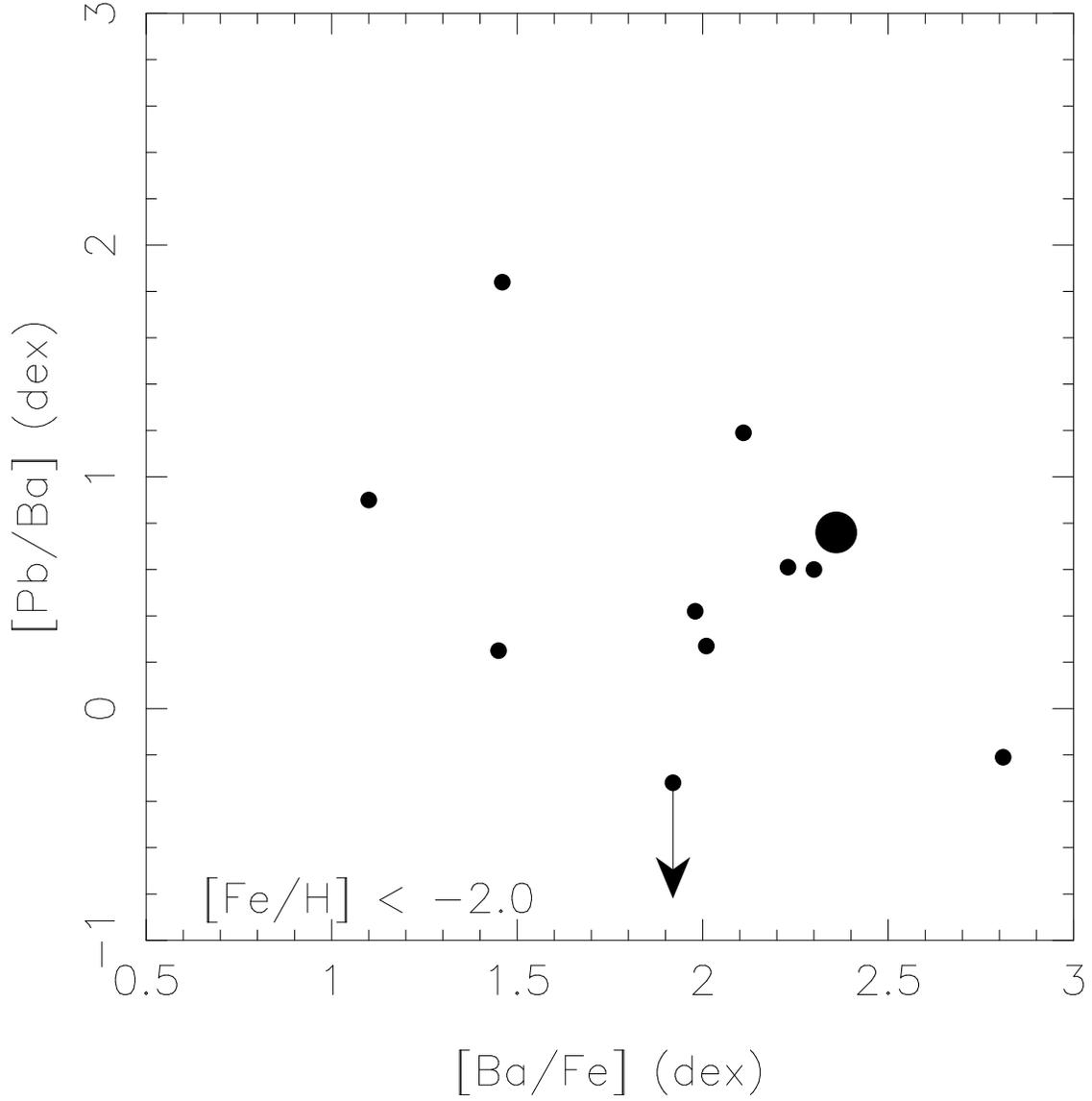}
\caption[]{The enhancement of lead as compared to Ba is plotted as a
function of the $s$-process enhancement (characterized by [Ba/Fe]) for very
metal poor ([Fe/H] $\le -2.0$ dex) 
$s$-process dominated stars.  The same sample shown in 
Figure~\ref{figure_senhance}, for which Pb abundances or upper limits 
could be located
in the literature, is used.  \mystar\ is indicated by the large filled circle.
\label{figure_bapb}}
\end{figure}

It appears that the relative amount of lead produced in
such $s$-process events is not constant.  While in some cases, including
that of \mystar\ and \sarastar\ \citep{lucatello03}, it
is as large as was predicted by \cite{gallino98} and
\cite{busso99}, in other cases the Pb enhancement is more modest.
As is widely recognized \citep*[see, e.g.][]{busso99}, 
the degree of Pb enhancement depends on the neutron to Fe 
seed ratio.
Figure~\ref{figure_bapb}
displays the [Pb/Ba] ratio as a function of the $s$-process enhancement
(parameterized in this case by [Ba/Fe]) for the same set of stars
shown in Figure~\ref{figure_senhance},  all stars with 
[Fe/H] $\le -2.0$ dex.
As noted by others, there is a considerable enhancement of Pb but well 
below the values to be expected if the neutron exposure were the same at 
low [Fe/H] as for the solar mainstream $s$ process. It follows that the 
neutron source strength cannot be the same as calculated for the main 
``solar'' $s$ process. Otherwise the conversion of Fe to Pb would be almost 
complete at [Fe/H]~$\sim -2.4$.  We infer that the neutron exposure roughly 
scales with [Fe/H] but with some variation producing higher Pb but not 
greatly altering the Ba peak.

The origin of the apparent $r$-process enhancement in \mystar\ is, however,
not known.  Such a large $r$-process enhancement is very rare.
The enhancement of Eu found among the ``pure'' $r$-process enhanced 
stars is much lower
(see Figure~\ref{figure_rscontrib}).
In previous studies it was argued that the high enrichments of 
$r$-process elements in a low-metallicity star must be the result of 
surface contamination by a binary companion subsequent to the formation
of the binary and do not represent the composition of the interstellar 
medium (ISM) from which the star was formed \citep{qian01}.
Hence one possible way to produce both $s$ and $r$ enhancement is to
consider that \mystar\ was the least massive member of a triple system.  
It could then have
obtained its $r$-process material from
a massive star in this system which became a Type II SN.
Then, assuming the binary did not become unbound by the SN explosion
nor the low mass star ablated away, 
the intermediate
mass component of this system, as it passed through the AGB phase, 
could have produced the $s$-process material and the excess of C and N, and transferred this material across the
binary to \mystar.  A triple star system with a history as described
above appears to be an implausible scenario.

Another possibility is that our understanding
of the $s$-process yields in AGB stars is flawed, and
that all of the Eu as well as the Ba observed in \mystar\
was produced through the $s$-process.  The range
of Ba/Eu predicted from AGB nucleosynthesis in different
calculations is fairly broad.  \cite{arlandini99} predict
Ba/Eu~640,  which is slightly smaller than that
of \cite{kappeler89} as quoted in \cite{burris00} for the solar
$s$-process inventory, 
but this is still far larger than the ratio observed in \mystar. 
However, the  predicted  Ba/Eu ratio for the 
most metal poor case (and only that particular case) 
of AGB nucleosynthesis by 
\cite{goriely00} of $\sim$105 is much closer to that
of \mystar\ than those of most other groups.  In either case, the
absolute abundances of the heavy elements predicted after dredge up at the
surface of a very metal poor AGB star are insufficient 
to explain the observations.
Furthermore, as is shown in Figure~\ref{figure_rscontrib},
very metal poor stars with [Fe/H] $-2.2$ to $-2.6$ with
enhanced abundances of the heavy neutron capture elements
which appear dominated by the $s$-process show Ba/Eu ratios
of 300 to 800, considerably larger than that of \mystar.  
Hence this hypothesis too is rejected.

The theoretical calculations of $s$-process production via
AGB nucleosynthesis, followed by
dredge up to the surface of the AGB star, then by mass
transfer across a binary system are very complex.  As reviewed
by \cite{busso99}, models of nucleosynthesis in AGB stars
have advanced greatly over the past
few years, particularly at near solar metallicity, although the problem
of galactic enrichment over a wide range of metallicity is not
adequately addressed.  
However, it is clear that changes in AGB models
to fit such an extreme case as \mystar\ seem to be required.

\subsection{Accretion Induced Collapse, Two Stars Instead of Three}

Accretion-induced collapse (AIC) of a white dwarf in a
binary is an  
interesting possibility to produce the enhancements seen in \mystar.
This produces a bare neutron star, which also might satisfy the 
requirement for the source of the heavy $r$-elements. The AIC mechanisms were
discussed earlier \citep*[e.g.][]{nomoto91}
and the possibility of AIC 
events being the source of $r$-elements was pointed out by 
\cite{woosley92}.  The case for AIC events being a source of the heavy 
$r$-elements will be presented in a companion paper, \cite{qian03}.
The data on \mystar\ were critical in this inference. The overall 
model is that a binary system is initially formed with one star being of 
mass $M\sim 3$--$12\,M_\odot$ and the other being of $M\lesssim 1\,M_\odot$.
The more massive star evolves through the AGB phase producing $s$-elements,
some of which are transferred to its lower-mass companion. Depending on
the ratio of neutrons to the Fe seeds in the AGB star, the $s$-process
may produce an extremely high Pb/Ba ratio or a pattern similar to the
solar main $s$-pattern. The more massive 
star eventually loses its envelope and becomes a white dwarf, 
which upon accreting material from the 
companion, subsequently collapses into a neutron star. 
The surface layers of the neutron star are heated by an enormous 
flux of neutrinos. This drives a wind, in which the heavy $r$-elements are 
produced. Some of the heavy $r$-elements are again transferred to the
companion. We consider this scenario to be a self-consistent explanation 
for the data on \mystar\ and for the observations of other 
low-metallicity stars with high enrichments ranging from predominantly 
$r$-process to predominantly $s$-process in origin. Further, we note that
the binary might be disrupted during the AIC event. This would explain
why a number of low-metallicity stars with high $s$-process enrichments
appear to be single stars \citep{preston01}.

\section{Comments on Cosmochronology}

Th/Eu cosmochronology depends critically on the assumption
that the initial production ratio of these two elements
is constant and is that given by the $r$-process. 
Th/Eu for \mystar\ is consistent with the expected (old) age for
very metal poor stars 
to within the (large) uncertainties.  The Th/Eu values for the bright stars
studied by \cite{johnson02} and for
BD+17$^{\circ}3248$ \citep{cowan02} are as well.  

However,  
CS~22892-052 has an apparent Th/Eu 
age of $\sim$16 Gyr while
CS~31082-001 has a very young Th/Eu age \cite{hill02}.  
Among the possible causes
of this are non-uniformities in the $r$-process yields
(see, for example, Otsuki, Matthews \& Kajino 2002) or 
an underestimate in the errors in the abundance determinations for the elements
Th and/or Eu in these two stars\footnote{An uncertainty
of 0.10 dex corresponds to an age uncertainty of $\sim$4 Gyr.}.
The first of the above difficulties, as well as the impact of potential
$s$-process contributions to the Eu abundance such as occurs
in \mystar, might be decreased considerably
by moving from a chronology based on Th/Eu to one
based on Th/U, but detection of U is quite difficult.

Our analysis of \mystar\ provides 
good evidence for mixing of $r$ and $s$-process enhancements, which
in turn leads to variation of the relative abundances of the heavy
neutron capture elements, in particular Ba/Eu and hence Th/Eu, 
in very metal poor stars.
We are still far from an understanding of the details of the
nucleosynthesis sites and production mechanisms for the heavy
elements in very metal poor stars, hence the use of Eu/Th
as a chronometer seems fraught with difficulties.

\acknowledgements
The entire Keck/HIRES user community owes a huge debt
to Jerry Nelson, Gerry Smith, Steve Vogt, and many other people who have
worked to make the Keck Telescope and HIRES a reality and to
operate and maintain the Keck Observatory.  
We are grateful to the W. M. Keck Foundation for the vision to 
fund the construction of the W. M. Keck Observatory.  
This publication makes use of data products from the Two Micron All Sky
Survey, which is a joint project of the University of Massachusetts and
the Infrared Processing and Analysis Center/California Institute
of Technology, funded by the National Aeronautics and Space Administration
and the National Science Foundation. 
We thank C.Conselice and I. Ivans for acquiring the
direct images used for photometry and
A.McWilliam for providing a code to
calculate hfs.
JGC is grateful for partial support from  NSF grants
AST-9819614 and AST-0205951.
N.C. acknowledges financial support through a Marie Curie Fellowship
of the European Community program \emph{Improving Human Research
Potential and the Socio-Economic Knowledge} under contract number
HPMF-CT-2002-01437, and from Deutsche Forschungsgemeinschaft under
grant Re 353/44-1.  Y.Z.Q.
is supported by DOE grants DE-FG02-87ER40328 
and DE-FG02-00ER41149 and G.J.W. by NASA grant 
NAG5-11725.  GJW is supported by NASA NAG5-11725. 
GPS contribution number 8896(1098).


\clearpage

\begin{deluxetable}{lcrrrr}
\tablenum{1}
\tablewidth{0pt}
\small
\tablecaption{\eqw\ for Lines in the Spectrum of \mystar\ \label{table_eqwidth}}
\tablehead{\colhead{Species} & \colhead{Wavelength} & \colhead{EP} 
& \colhead{Log($gf$)]} & \colhead{$C_6$} & \colhead{\eqw} \\
\colhead{} & \colhead{(\AA)} & \colhead{(eV)} & \colhead{(dex)} 
& \colhead{(Hz cm$^6$)\tablenotemark{a}} & \colhead{(m\AA)}  }
\startdata 
 Mg~I   &  4057.52 &   4.34 &  $-$1.200 &  0.630E$-$29 &    14.0    \\ 
 Mg~I   &  4167.28 &   4.34 &  $-$1.000 &  0.630E$-$29 &    39.5    \\ 
 Mg~I   &  4703.00 &   4.34 &  $-$0.670 &  0.132E$-$29 &    46.3    \\ 
 Mg~I   &  5172.70 &   2.71 &  $-$0.380 &  0.138E$-$30 &   130.7    \\ 
 Mg~I   &  5183.62 &   2.72 &  $-$0.160 &  0.138E$-$30 &   159.6    \\ 
 Al~I   &  3961.52 &   0.00 &  $-$0.340 &  0.631E$-$30 &    78.4    \\ 
 Si~I   &  3905.53 &   1.91 &  $-$1.000 &  0.190E$-$30 &   149.5    \\ 
 Ca~I   &  4226.74 &   0.00 &   0.240 &  0.589E$-$31 &   137.2    \\ 
 Ca~I   &  4318.66 &   1.90 &  $-$0.210 &  0.589E$-$30 &    29.7    \\ 
 Ca~I   &  4425.44 &   1.88 &  $-$0.360 &  0.589E$-$30 &    26.6    \\ 
 Ca~I   &  4435.69 &   1.89 &  $-$0.520 &  0.589E$-$30 &    36.7    \\ 
 Ca~I   &  4454.79 &   1.90 &   0.260 &  0.589E$-$30 &    61.0    \\ 
 Sc~II  &  4246.82 &   0.32 &   0.242 &     2.0   &    79.2    \\ 
 Sc~II  &  4320.73 &   0.60 &  $-$0.260 &     2.0   &    36.6    \\ 
 Sc~II  &  4670.41 &   1.36 &  $-$0.580 &     2.0   &    14.6    \\ 
 Ti~I   &  4533.25 &   0.85 &   0.480 &     2.0   &    14.0    \\ 
 Ti~I   &  4981.74 &   0.85 &   0.500 &     2.0   &    17.5    \\ 
 Ti~I   &  4999.51 &   0.83 &   0.250 &     2.0   &    18.0    \\ 
 Ti~II  &  3900.54 &   1.13 &  $-$0.450 &     2.0   &    81.6    \\ 
 Ti~II  &  4028.35 &   1.89 &  $-$0.870 &     2.0   &    30.9    \\ 
 Ti~II  &  4395.03 &   1.08 &  $-$0.510 &     2.0   &    87.1    \\ 
 Ti~II  &  4399.77 &   1.24 &  $-$1.300 &     2.0   &    48.6    \\ 
 Ti~II  &  4417.72 &   1.16 &  $-$1.200 &     2.0   &    38.4    \\ 
 Ti~II  &  4443.81 &   1.08 &  $-$0.700 &     2.0   &    73.0    \\ 
 Ti~II  &  4468.51 &   1.13 &  $-$0.600 &     2.0   &    75.1    \\ 
 Ti~II  &  4501.28 &   1.12 &  $-$0.760 &     2.0   &    67.8    \\ 
 Ti~II  &  4533.97 &   1.24 &  $-$0.640 &     2.0   &    67.9    \\ 
 Ti~II  &  4563.77 &   1.22 &  $-$0.820 &     2.0   &    58.5    \\ 
 Ti~II  &  4571.98 &   1.57 &  $-$0.340 &     2.0   &    66.8    \\ 
 Ti~II  &  4589.95 &   1.24 &  $-$1.600 &     2.0   &    16.5    \\ 
 Ti~II  &  5185.91 &   1.89 &  $-$1.500 &     2.0   &     6.0    \\ 
 Cr~I   &  4254.33 &   0.00 &  $-$0.110 &     2.0   &    52.3    \\ 
 Cr~I   &  4274.79 &   0.00 &  $-$0.230 &     2.0   &    48.2    \\ 
 Cr~I   &  5206.04 &   0.94 &   0.030 &     2.0   &    21.3    \\ 
 Mn~I   &  4030.75 &   0.00 &  $-$0.470 &     2.0   &    34.5    \\ 
 Mn~I   &  4033.06 &   0.00 &  $-$0.620 &     2.0   &    31.1    \\ 
 Fe~I   &  3865.52 &   1.01 &  $-$0.980 &     2.0   &    68.3    \\ 
 Fe~I   &  3899.72 &   0.09 &  $-$1.500 &     2.0   &    82.8    \\ 
 Fe~I   &  3902.96 &   1.56 &  $-$0.470 &     2.0   &    70.4    \\ 
 Fe~I   &  3920.27 &   0.12 &  $-$1.800 &     2.0   &    70.0    \\ 
 Fe~I   &  3922.92 &   0.05 &  $-$1.600 &     2.0   &    83.8    \\ 
 Fe~I   &  3949.96 &   2.18 &  $-$1.200 &     2.0   &    12.8    \\ 
 Fe~I   &  4005.24 &   1.56 &  $-$0.610 &     2.0   &    70.8    \\ 
 Fe~I   &  4045.81 &   1.49 &   0.280 &     2.0   &   107.7    \\ 
 Fe~I   &  4063.59 &   1.56 &   0.060 &     2.0   &    94.0    \\ 
 Fe~I   &  4071.74 &   1.61 &  $-$0.020 &     2.0   &    93.2    \\ 
 Fe~I   &  4118.55 &   3.57 &   0.140 &     2.0   &    33.6    \\ 
 Fe~I   &  4132.06 &   1.61 &  $-$0.820 &     2.0   &    64.7    \\ 
 Fe~I   &  4143.87 &   1.56 &  $-$0.620 &     2.0   &    74.6    \\ 
 Fe~I   &  4147.67 &   1.49 &  $-$2.100 &     2.0   &    12.0    \\ 
 Fe~I   &  4181.75 &   2.83 &  $-$0.370 &     2.0   &    38.6    \\ 
 Fe~I   &  4187.05 &   2.45 &  $-$0.550 &     2.0   &    42.0    \\ 
 Fe~I   &  4187.81 &   2.43 &  $-$0.550 &     2.0   &    48.0    \\ 
 Fe~I   &  4198.33 &   2.40 &  $-$0.720 &     2.0   &    55.0    \\ 
 Fe~I   &  4199.10 &   3.05 &   0.160 &     2.0   &    52.6    \\ 
 Fe~I   &  4202.04 &   1.49 &  $-$0.710 &     2.0   &    65.2    \\ 
 Fe~I   &  4216.19 &   0.00 &  $-$3.400 &     2.0   &    13.0    \\ 
 Fe~I   &  4222.22 &   2.45 &  $-$0.970 &     2.0   &    10.1    \\ 
 Fe~I   &  4227.44 &   3.33 &   0.270 &     2.0   &    41.3    \\ 
 Fe~I   &  4233.61 &   2.48 &  $-$0.600 &     2.0   &    35.7    \\ 
 Fe~I   &  4250.13 &   2.47 &  $-$0.410 &     2.0   &    39.9    \\ 
 Fe~I   &  4250.80 &   1.56 &  $-$0.380 &     2.0   &    65.5    \\ 
 Fe~I   &  4260.49 &   2.40 &   0.140 &     2.0   &    60.9    \\ 
 Fe~I   &  4271.16 &   2.45 &  $-$0.350 &     2.0   &    63.7    \\ 
 Fe~I   &  4271.77 &   1.49 &  $-$0.160 &     2.0   &    85.6    \\ 
 Fe~I   &  4282.41 &   2.18 &  $-$0.780 &     2.0   &    41.8    \\ 
 Fe~I   &  4294.14 &   1.49 &  $-$0.970 &     2.0   &    77.1    \\ 
 Fe~I   &  4325.77 &   1.61 &   0.010 &     2.0   &    85.2    \\ 
 Fe~I   &  4375.94 &   0.00 &  $-$3.000 &     2.0   &    24.6    \\ 
 Fe~I   &  4383.56 &   1.49 &   0.200 &     2.0   &   104.8    \\ 
 Fe~I   &  4404.76 &   1.56 &  $-$0.140 &     2.0   &    88.6    \\ 
 Fe~I   &  4415.13 &   1.61 &  $-$0.610 &     2.0   &    73.8    \\ 
 Fe~I   &  4427.32 &   0.05 &  $-$3.000 &     2.0   &    38.3    \\ 
 Fe~I   &  4442.35 &   2.20 &  $-$1.200 &     2.0   &    36.4    \\ 
 Fe~I   &  4459.14 &   2.18 &  $-$1.300 &     2.0   &    23.0    \\ 
 Fe~I   &  4461.66 &   0.09 &  $-$3.200 &     2.0   &    17.7    \\ 
 Fe~I   &  4482.17 &   0.11 &  $-$3.500 &     2.0   &    18.6    \\ 
 Fe~I   &  4494.57 &   2.20 &  $-$1.100 &     2.0   &    18.0    \\ 
 Fe~I   &  4531.16 &   1.49 &  $-$2.200 &     2.0   &    10.9    \\ 
 Fe~I   &  4592.66 &   1.56 &  $-$2.500 &     2.0   &    10.0    \\ 
 Fe~I   &  4602.95 &   1.49 &  $-$2.200 &     2.0   &    11.0    \\ 
 Fe~I   &  4654.50 &   1.56 &  $-$2.800 &     2.0   &     5.3    \\ 
 Fe~I   &  4871.33 &   2.86 &  $-$0.360 &     2.0   &    35.5    \\ 
 Fe~I   &  4872.14 &   2.88 &  $-$0.570 &     2.0   &    29.9    \\ 
 Fe~I   &  4891.50 &   2.85 &  $-$0.110 &     2.0   &    37.1    \\ 
 Fe~I   &  4919.00 &   2.86 &  $-$0.340 &     2.0   &    28.7    \\ 
 Fe~I   &  4920.51 &   2.83 &   0.150 &     2.0   &    60.0    \\ 
 Fe~I   &  4957.61 &   2.81 &   0.230 &     2.0   &    70.9    \\ 
 Fe~I   &  5171.61 &   1.48 &  $-$1.800 &     2.0   &    25.6    \\ 
 Fe~I   &  5194.95 &   1.56 &  $-$2.100 &     2.0   &    10.8    \\ 
 Fe~I   &  5216.28 &   1.61 &  $-$2.200 &     2.0   &     8.6    \\ 
 Fe~I   &  5232.95 &   2.94 &  $-$0.100 &     2.0   &    38.0    \\ 
 Fe~I   &  5269.55 &   0.86 &  $-$1.300 &     2.0   &    72.8    \\ 
 Fe~I   &  5324.19 &   3.21 &  $-$0.100 &     2.0   &    21.8    \\ 
 Fe~II  &  4178.86 &   2.57 &  $-$2.500 &     2.0   &    16.8    \\ 
 Fe~II  &  4416.82 &   2.77 &  $-$2.400 &     2.0   &    16.0    \\ 
 Fe~II  &  4491.40 &   2.84 &  $-$2.600 &     2.0   &     6.1    \\ 
 Fe~II  &  4508.30 &   2.84 &  $-$2.300 &     2.0   &    20.0    \\ 
 Fe~II  &  4555.89 &   2.82 &  $-$2.200 &     2.0   &    21.6    \\ 
 Fe~II  &  4583.84 &   2.81 &  $-$2.000 &     2.0   &    45.0    \\ 
 Fe~II  &  4923.93 &   2.88 &  $-$1.300 &     2.0   &    57.8    \\ 
 Fe~II  &  5018.45 &   2.89 &  $-$1.200 &     2.0   &    69.8    \\ 
 Fe~II  &  5197.58 &   3.23 &  $-$2.200 &     2.0   &    12.4    \\ 
 Fe~II  &  5234.63 &   3.22 &  $-$2.200 &     2.0   &    16.2    \\ 
 Co~I   &  3845.46 &   0.92 &   0.009 &     2.0   &    24.8    \\ 
 Co~I   &  3873.11 &   0.43 &  $-$0.666 &     2.0   &    25.5    \\ 
 Co~I   &  4121.31 &   0.92 &  $-$0.315 &     2.0   &    10.6    \\ 
 Ni~I   &  3858.30 &   0.42 &  $-$0.967 &     2.0   &    65.1    \\ 
 Ni~I   &  4401.55 &   3.19 &   0.084 &     2.0   &     9.5    \\ 
 Sr~II  &  4077.71 &   0.00 &   0.170 &  0.254E$-$31 &   125.1    \\ 
 Sr~II  &  4215.52 &   0.00 &  $-$0.140 &  0.254E$-$31 &   110.3    \\
 Y~II   &  3818.34 &   0.13 &  $-$0.980 &     2.0   &    11.7    \\ 
 Y~II   &  3950.36 &   0.10 &  $-$0.490 &     2.0   &    25.9    \\ 
 Y~II   &  4398.01 &   0.13 &  $-$1.000 &     2.0   &    16.1    \\ 
 Y~II   &  4883.69 &   1.08 &   0.070 &     2.0   &    14.7    \\ 
 Y~II   &  5087.43 &   1.08 &  $-$0.170 &     2.0   &    10.0    \\ 
 Zr~II  &  4161.21 &   0.71 &  $-$0.720 &     2.0   &    21.6    \\ 
 Zr~II  &  4208.99 &   0.71 &  $-$0.460 &     2.0   &    20.9    \\ 
 Zr~II  &  4317.32 &   0.71 &  $-$1.400 &     2.0   &     6.2    \\ 
 Ba~II  &  4130.65 &   2.72 &   0.560 &  0.500E$-$31 &    55.0    \\ 
 Ba~II  &  4554.04 &   0.00 &   0.170 &  0.224E$-$31 &   195.4    \\ 
 Ba~II  &  4934.16 &   0.00 &  $-$0.150 &  0.357E$-$31 &   190.4    \\ 
 La~II  &  3988.52 &   0.40 &   0.210 &     2.0   &    65.4    \\ 
 La~II  &  3995.75 &   0.17 &  $-$0.060 &     2.0   &    58.1    \\ 
 La~II  &  4086.71 &   0.00 &  $-$0.070 &     2.0   &    55.1    \\ 
 La~II  &  4123.23 &   0.32 &   0.130 &     2.0   &    45.4    \\ 
 La~II  &  4333.76 &   0.17 &  $-$0.060 &     2.0   &    84.5    \\ 
 La~II  &  5122.99 &   0.32 &  $-$0.850 &     2.0   &    15.0    \\ 
 Ce~II  &  4073.47 &   0.48 &   0.320 &     2.0   &    21.9    \\ 
 Ce~II  &  4083.23 &   0.70 &   0.240 &     2.0   &    23.0    \\ 
 Ce~II  &  4120.84 &   0.32 &  $-$0.240 &     2.0   &    14.8    \\ 
 Ce~II  &  4127.38 &   0.68 &   0.240 &     2.0   &    22.5    \\ 
 Ce~II  &  4418.79 &   0.86 &   0.310 &     2.0   &    34.9    \\ 
 Ce~II  &  4486.91 &   0.30 &  $-$0.360 &     2.0   &    17.2    \\ 
 Ce~II  &  4562.37 &   0.48 &   0.330 &     2.0   &    28.5    \\ 
 Pr~II  &  3964.26 &   0.22 &  $-$0.400 &     2.0   &    10.4    \\ 
 Pr~II  &  4062.80 &   0.42 &   0.330 &     2.0   &    13.8    \\ 
 Pr~II  &  5220.12 &   0.80 &   0.170 &     2.0   &     4.1    \\ 
 Nd~II  &  4018.81 &   0.06 &  $-$0.880 &     2.0   &     6.8    \\ 
 Nd~II  &  4021.34 &   0.32 &  $-$0.170 &     2.0   &    20.9    \\ 
 Nd~II  &  4061.09 &   0.47 &   0.300 &     2.0   &    41.5    \\ 
 Nd~II  &  4069.27 &   0.06 &  $-$0.400 &     2.0   &    16.5    \\ 
 Nd~II  &  4109.46 &   0.32 &   0.180 &     2.0   &    56.9    \\ 
 Nd~II  &  4232.38 &   0.06 &  $-$0.350 &     2.0   &    16.7    \\ 
 Nd~II  &  4446.39 &   0.20 &  $-$0.630 &     2.0   &    20.6    \\ 
 Nd~II  &  4462.99 &   0.56 &  $-$0.070 &     2.0   &    20.3    \\ 
 Nd~II  &  5212.35 &   0.20 &  $-$0.700 &     2.0   &     9.4    \\ 
 Sm~II  &  4318.94 &   0.28 &  $-$0.270 &     2.0   &     7.9    \\ 
 Sm~II  &  4537.95 &   0.48 &  $-$0.230 &     2.0   &     8.3    \\ 
 Eu~II  &  3819.67 &   0.00 &   0.510 &     2.0   &    68.4    \\ 
 Eu~II  &  3907.11 &   0.21 &   0.170 &     2.0   &    24.1    \\ 
 Eu~II  &  4129.70 &   0.00 &   0.220 &     2.0   &    58.3    \\ 
 Gd~II  &  3844.58 &   0.14 &  $-$0.510 &     2.0   &    12.7    \\ 
 Gd~II  &  3916.51 &   0.60 &   0.060 &     2.0   &    12.6    \\ 
 Gd~II  &  4037.91 &   0.73 &   0.070 &     2.0   &    10.0    \\ 
 Gd~II  &  4191.08 &   0.43 &  $-$0.680 &     2.0   &     5.2    \\ 
 Dy~II  &  3869.86 &   0.00 &  $-$0.940 &     2.0   &    14.9    \\ 
 Dy~II  &  3996.69 &   0.59 &  $-$0.190 &     2.0   &     8.0    \\ 
 Dy~II  &  4103.31 &   0.10 &  $-$0.370 &     2.0   &    25.0    \\ 
\tablenotetext{a}{This is either the value of $C_6$
for van der Waals broadening or the multiplier
to be applied to the  Uns\"{o}ld (1955) approximation for $C_6$.} 
\enddata
\end{deluxetable}

\begin{deluxetable}{lrrrr}
\tablenum{2}
\tablewidth{0pt}
\small
\tablecaption{Radial Velocity Monitoring \label{table_vr}}
\tablehead{\colhead{Star} & \colhead{HJD\tablenotemark{a}} &  
\colhead{$v_r$(Helio)} & \colhead{$\sigma$} & \colhead{Exp. Time} \\
\colhead{} & \colhead{} & \colhead{(\kms)} & \colhead{(\kms)} 
& \colhead{(sec)} }
\startdata 
HE~0007$-$1832 & 181.93  & +27.3 & 0.8 & 4800 \\
HE~0007$-$1832 & 182.87  & +28.2 & 0.8 & 3600  \\
HE~0007$-$1832 & 545.93   & +26.4 & 0.9 & 2400 \\
~  \\
\mystar & 180.72 & +41.3 & 0.5 & 7200 \\
\mystar & 182.77 & +41.6 & 0.4 & 6000 \\
\mystar & 397.12 & +38.3 & 0.4 & 1200 \\
\mystar & 544.78 & +36.8 & 0.5 & 3600 \\
~ \\
HD~186478 & 179.71 & +25.7 & 0.6 & 20 \\
HD~186478 & 544.77 & +31.2 & 0.5 & 50 \\
HD~186478 & 545.76 & +31.0 & 0.5 & 50 \\
HD~186478 & 546.76 & +31.3 & 0.6 & 50 \\
\tablenotetext{a}{Heliocentric Julian date - 2452000.00.}
\enddata
\end{deluxetable}

%
\begin{deluxetable}{lcllrr}
\tablenum{3}
\tablewidth{0pt}
\small
\tablecaption{Abundances in \mystar \label{table_abund}}
\tablehead{\colhead{Species} & \colhead{No. Lines} &  
\colhead{Log[$\epsilon(X)$]\tablenotemark{a}} & 
\colhead{$\Delta$(non-LTE)\tablenotemark{b}} &
\colhead{$<$[X/Fe]$>$} &
\colhead{Solar Log[$\epsilon(X)$]\tablenotemark{c}} \\
\colhead{} & \colhead{} & \colhead{(dex)} & \colhead{(dex)} &
\colhead{(dex)} & \colhead{(dex)}  }
\startdata 
C (CH)\tablenotemark{d}  & syn &    8.08 (0.2) & \nodata & 1.91 & 8.56  \\
N (CN) & syn & 7.25 (0.3) & \nodata & 1.65  &  7.92 \\
Mg I    &  5  &   5.72 (0.25) &  +0.15 & 0.50 &     7.54 \\
Al I    &  1  &   3.39 (...) & +0.6\tablenotemark{e} & $-$0.76 &     6.47 \\
Si I    &  1  &   5.92 (...) & \nodata & 0.69 &     7.55   \\
Ca I    &  5  &   4.49 (0.21) & \nodata & 0.45 &     6.36 \\
Sc II   &  3  &   1.42 (0.24) & \nodata & 0.59 &  4.99  \\
Ti I    &  3  &   3.22  (0.17) & \nodata & 0.55 &     4.99 \\  
Ti II   &  13 &   3.20 (0.13)  & \nodata & 0.53 &     4.99 \\
Cr I    &  3  &   3.01 (0.02)  & \nodata & $-$0.35 &     5.67 \\ 
Mn I    &  2  &   2.46 (0.06) & \nodata &  $-$0.61 &  5.39 \\
Fe I    &  58 &   5.13 (0.20) & +0.10 & 0.00 &     7.45 \\
Fe II   &  10 &   5.17 (0.14)  &  0.00 & 0.04 &     7.45 \\
Co I    &   3 &   2.81 (0.20)  & \nodata & 0.21 &  4.92  \\
Ni I    &   2 &   3.99 (0.10)  & \nodata & 0.06 &     6.25 \\
Sr II   &   2 &   1.34 (0.02)  & \nodata &  0.76 &     2.90 \\
Y II    &   5 &   0.75 (0.12)  & \nodata & 0.83 &     2.24 \\ 
Zr II   &   3 &   1.75 (0.18)  & \nodata & 1.47 &     2.60 \\
Ba II   &   3 &   2.17 (0.07)  & +0.15 & 2.36 &     2.13 \\
La II   &   6 &   1.10 (0.20)  & \nodata & 2.38 &     1.14 \\
Ce II   &   7 &   1.51 (0.20)  & \nodata & 2.28 &     1.55 \\
Pr II   &   3 &   0.91 (0.30)  & \nodata & 2.52 &     0.71 \\
Nd II   &   9 &   1.45 (0.22)  & \nodata & 2.27 &     1.50 \\
Sm II   &   2 &   0.67 (0.11)  & \nodata & 1.99 &     1.00 \\
Eu II   &   3 &   0.17 (0.10)  & $-$0.10 & 1.98 &     0.51 \\
Gd II   &   4 &   1.04 (0.10)  & \nodata & 2.23 &     1.12 \\
Dy II   &   3 &   0.95 (0.30)  & \nodata & 2.16 &     1.10 \\
Pb I    &   syn & 2.8 (0.2) &  \nodata &  3.12 &     2.00 \\
Th II   &  syn &  $-$0.5 (+0.2,$-\infty$)\tablenotemark{f} & \nodata &  1.73  &   0.09 \\
\tablenotetext{a}{The 1$\sigma$ errors for log($\epsilon$) are given in
parentheses.}
\tablenotetext{b}{The non-LTE corrections are derived from the calculations
referenced in \S\ref{section_nonlte}.}
\tablenotetext{c}{The Solar abundances we adopt
are described in \S\ref{section_sun}.}
\tablenotetext{d}{Synthesis of the CH band shows \ciso\ $\sim$10; see text.}
\tablenotetext{e}{The non-LTE correction for the resonance doublet
of Al~I is used throughout this paper.}
\tablenotetext{f}{The detection of Th is not secure.  See the text for details.}
\enddata
\end{deluxetable}

\clearpage

\begin{deluxetable}{lcrrrr}
\tablenum{4}
\tablewidth{0pt}
\small
\tablecaption{Sensitivity of Abundances \label{sens_table}}
\tablehead{\colhead{Species} & 
\colhead{$\Delta(T_{eff}=$} & 
\colhead{$\Delta$(log($g$)} & 
\colhead{$\Delta$[Fe/H](model)}
& \colhead{$\Delta(v_t= -0.2$} & 
\colhead{$\Delta$[log($C_6)=$} \\
\colhead{} & \colhead{-100K) (dex)\tablenotemark{a}} 
& \colhead{=+0.2 dex) (dex)\tablenotemark{a}} 
& \colhead{=+0.5 dex) (dex)\tablenotemark{a}} &
\colhead{\kms) (dex)\tablenotemark{a}} 
& \colhead{+0.3 dex] (dex)\tablenotemark{a}} }
\startdata 
C (CH)  &  $-$0.12 & $-$0.08 & $-0.13$ & 0.04 & \nodata \\
N (CN)  &  $-$0.08 & $-$0.06 & $-0.60$\tablenotemark{b} & 0.01 & \nodata \\
Mg I    &  $-$0.06 & 0.01 & 0.01 & 0.03 & $-$0.03 \\
Al I    &  $-$0.08 & 0.01 & 0.02 & 0.06 &  $-$0.03\\
Si I    &  $-$0.11 & 0.04 & 0.02 & 0.05 & $-$0.08  \\
Ca I    &  $-$0.07 & 0.01 & 0.02 & 0.03 & $-0.02$ \\
Sc II   &  $-0.04$ & $-0.02$ & 0.01 & 0.05\tablenotemark{c} & \nodata  \\
Ti I    &  $-0.08$ & 0.00 & 0.02 & 0.01 & \nodata\\  
Ti II   &  $-0.04$ & $-$0.03 & 0.00 & 0.06 & \nodata \\
Cr I    &  $-0.09$ & 0.00 & 0.02 & 0.03 & \nodata \\ 
Mn I    &  $-$0.10 & 0.00 & 0.02 & 0.02\tablenotemark{c} & \nodata \\
Fe I    &  $-0.08$ & 0.00 & 0.01 & 0.05 & \nodata \\
Fe II   &  $-$0.02 & $-$0.03 & 0.00 & 0.02 & \nodata\\
Co I    &  $-0.09$ & 0.00 & 0.02 & 0.01\tablenotemark{c} & \nodata \\
Ni I    &  $-0.08$ & 0.00 & 0.02 & 0.05 & \nodata \\
Sr II   &  $-0.09$ & $-0.01$ & $-$0.01 & 0.11 & $-0.06$ \\
Y II    &  $-0.05$ & $-0.03$ & 0.02 & 0.01 & \nodata \\ 
Zr II   &  $-0.05$ & $-0.03$ & 0.02 & 0.01 & \nodata \\
Ba II   &  $-0.08$ & 0.00 & 0.01 & $-0.01$\tablenotemark{c} 
& $-0.06$\tablenotemark{c} \\
La II   &  $-0.06$ & $-0.03$ & 0.01 & $-0.03$\tablenotemark{c} & \nodata \\
Ce II   &  $-0.05$ & $-0.03$ & 0.02 & 0.02 & \nodata \\
Pr II   &  $-0.06$ & $-0.03$ & 0.01 & 0.01 & \nodata \\
Nd II   &  $-0.06$ & $-0.03$  & 0.01 & 0.02 & \nodata \\
Sm II   &  $-0.06$ & $-0.03$ & 0.02 & 0.01 & \nodata \\
Eu II   &  $-0.06$ & $-0.03$ & 0.01 & $-0.01$\tablenotemark{c} & \nodata \\
Gd II   &  $-0.05$ & $-0.03$ & 0.01 & 0.01 & \nodata \\
Dy II   &  $-0.06$ & $-0.03$ & 0.01 & 0.02 & \nodata \\
Pb I    &  $-0.08$ &  0.00 & 0.00 & $-0.07$ & \nodata \\
Th II   &  $-0.06$ & $-0.03$ & 0.02 & 0.01 & \nodata \\
\enddata
\tablenotetext{a}{These are the changes in log[$\epsilon(X)]$, not
in [X/Fe].  To calculate the latter, subtract $\Delta$(Fe) from $\Delta$(X)
for a specific parameter, i.e. \teff, \grav, etc.}
\tablenotetext{b}{This value for CN is calculated for constant C/Fe.}
\tablenotetext{c}{Hyper fine structure corrections have 
been applied in each case.}
\end{deluxetable}

\clearpage

\begin{deluxetable}{lrrrr}
\tablenum{5}
\tablewidth{0pt}
\small
\tablecaption{Comparison of the Observed and Predicted Abundances in \mystar\
Using the Solar $r$ and $s$-Process Inventory
\label{table_r_s_sum}}
\tablehead{\colhead{Element} & \colhead{$\epsilon$(X)} &
\colhead{Pred. $\epsilon$(X--$r$)\tablenotemark{a}} &
\colhead{Pred. $\epsilon$(X--$s$)\tablenotemark{b}} &
\colhead{$\Delta$(Pred.-Obs.)} \\
\colhead{} & \colhead{(H=10$^{12}$)} & \colhead{(H=10$^{12}$)} &
\colhead{(H=10$^{12}$)} & \colhead{(dex)} }
\startdata 
 38 (Sr) &  21.78 &    56.10 &  603.72 &    1.48 \\
 39 (Y)  &  5.62 &   28.82 &  98.36 &    1.35 \\
 40 (Zr) & 56.23  &  44.88 &  255.03  &   0.73 \\
  56 (Ba) & 147.91 &     17.82  &  137.24  &    0.02 \\
 57 (La) &  12.59 &        2.42 &   10.01  &  $-$0.01 \\
  58 (Ce) &    32.21   &  4.40 &   26.21  &  $-$0.02 \\
 59 (Pr) &     8.09  &    1.80  &   2.30 &   $-$0.30 \\
 60 (Nd) &  28.38 &    9.24  &  11.19 &   $-$0.14 \\
 62 (Sm) &   4.71  &    3.74  &   2.65   &   0.13 \\
 63 (Eu) &  1.48 &  2.00  &   0.07  &   0.15 \\
 64 (Gd) &  10.86 &    6.17   &  1.80  &  $-$0.14 \\
 66 (Dy) &   9.04  &  7.92 &    1.47   &  0.02 \\
 82 (Pb) &   630.96 &   13.64  &  67.73 &   $-$0.89 \\ 
 90 (Th) &  0.32  &   0.92 &  \nodata  &   0.47 \\
\tablenotetext{a}{The $r$-process contribution in \mystar\ is set as 
0.62 times the
Solar $r$-process inventory given by \cite{burris00} in the same units.}
\tablenotetext{b}{The $s$-process contribution in \mystar\ is set as
the 0.83 times the
Solar $s$-process inventory given by \cite{burris00} in the same units.}
\enddata
\end{deluxetable}

\clearpage

\end{document}